\documentclass[showpacs,prl,onecolumn,aps,superscriptaddress,preprintnumbers,nofootinbib]{revtex4}
\usepackage[T1]{fontenc}
\usepackage[latin9]{inputenc}
\usepackage{amsmath,amssymb}
\usepackage{epsfig}
\usepackage{graphicx}
\usepackage{amsmath}
\usepackage{amsfonts}
\def\slashchar#1{\setbox0=\hbox{$#1$}     		% set a box for #1
   \dimen0=\wd0                                 	% and get its size
   \setbox1=\hbox{/} \dimen1=\wd1               	% get size of /
   \ifdim\dimen0>\dimen1                        	% #1 is bigger
      \rlap{\hbox to \dimen0{\hfil/\hfil}}      	% so center / in box
      #1                                        	% and print #1
   \else                                        	% / is bigger
      \rlap{\hbox to \dimen1{\hfil$#1$\hfil}}   	% so center #1
      /                                         	% and print /
   \fi}

\renewcommand{\vec}{\boldsymbol}
\newcommand{\beq}{\begin{equation}}
\newcommand{\eeq}{\end{equation}}
\newcommand{\bea}{\begin{eqnarray}}
\newcommand{\eea}{\end{eqnarray}}
\newcommand{\baa}{\begin{array}}
\newcommand{\eaa}{\end{array}}

\def\eq#1{{Eq.~(\ref{#1})}}

\def\fig#1{{Fig.~\ref{#1}}}
\newcommand{\aal}{\bar{\alpha}_S}
\newcommand{\bas}{\bar{\alpha}_S}
\newcommand{\as}{\alpha_S}

\newcommand{\nn}{\nonumber}
\newcommand{\ea}{e^{(\gamma-1)\xi}}

\newcommand{\D}{\partial}
\newcommand{\h}{\frac{1}{2}}

\newcommand{\x}{\vec{x}}

\newcommand{\ga}{\gamma}

\newcommand{\om}{\omega}

\newcommand{\Lb}{\left(}
\newcommand{\Rb}{\right)}
\def\pom{{I\!\!P}}

\newcommand{\dy}{\delta Y}
\newcommand{\pp}{\partial}

\renewcommand{\vec}[1]{\boldsymbol{#1}}

\newcommand{\gga}{\tilde{\gamma}}

\newcommand{\intee}{\int_{C_{3}}  \frac{d\gamma}{2\pi i} }
\begin{document}
%%%%%%%%%%%%%%%%%%%%%%%%%%%%%%%%%%%%%%%%%%%%%%%%%%%%%%%%%%%%%%%%%%%%%%%
\title{CGC/saturation approach: an impact-parameter dependent model for diffraction production in DIS.}
\author{Carlos Contreras}
\email{carlos.contreras@usm.cl}
\affiliation{Departemento de F\'isica, Universidad T\'ecnica Federico Santa Mar\'ia, and Centro Cient\'ifico-\\
Tecnol\'ogico de Valpara\'iso, Avda. Espana 1680, Casilla 110-V, Valpara\'iso, Chile}
\author{ Eugene ~ Levin}
\email{leving@post.tau.ac.il, eugeny.levin@usm.cl}
\affiliation{Departemento de F\'isica, Universidad T\'ecnica Federico Santa Mar\'ia, and Centro Cient\'ifico-\\
Tecnol\'ogico de Valpara\'iso, Avda. Espana 1680, Casilla 110-V, Valpara\'iso, Chile}
\affiliation{Department of Particle Physics, School of Physics and Astronomy,
Raymond and Beverly Sackler
 Faculty of Exact Science, Tel Aviv University, Tel Aviv, 69978, Israel}
\author{Rodrigo Meneses}
\email{rodrigo.meneses@uv.cl}
\affiliation{Escuela de Ingenier\'\i a Civil, Facultad de Ingenier\'\i a, Universidad de Valpara\'\i so, Avda Errazuriz 1834, Valpara\'\i so, Chile}
\author{  Irina Potashnikova}
\email{irina.potashnikova@usm.cl}
\affiliation{Departemento de F\'isica, Universidad T\'ecnica Federico Santa Mar\'ia, and Centro Cient\'ifico-\\
Tecnol\'ogico de Valpara\'iso, Avda. Espana 1680, Casilla 110-V, Valpara\'iso, Chile}
\date{\today}

%%%%%%%%%%%%%%%%%%%%%%%%%%%%%%%%%%%%%%%%%%%%%%%%%%%%%%%%%%%%%%%%
\begin{abstract}
 
  In the paper we discussed the evolution equations for  diffractive
 production in the framework of CGC/saturation approach,  and found the 
analytical solutions for several kinematic regions. 
  The most impressive features of these  solutions are,  that  
diffractive
 production does not exibit  geometric scaling behaviour i.e. being 
a function of
 one variable. 
  
  Based on these solutions, we suggest an impact parameter dependent 
saturation
 model, which is suitable for  describing  diffraction production 
both deep
 in the saturation region, and in the vicinity of the saturation scale.

  Using the model we attempted to fit the combined data on 
diffraction production 
from H1 and ZEUS collaborations. We found that we are able describe both
 $x_\pom$ and $\beta$ dependence, as well as $Q$ behavior of the measured
 cross sections.  In spite of the sufficiently large $\chi^2/d.o.f.$ we
 believe that our description  provides an  initial impetus  to 
find 
a fit
 of the experimental data, based on the solution of the CGC/saturation 
equation, rather than on describing the diffraction system in simplistic
 manner, assuming that only quark-antiquark pair and one extra 
gluons, are
 produced.

  \end{abstract}

\maketitle

\vspace{-0.5cm}
\tableofcontents

%\flushbottom

%\pagestyle{empty}

%\mbox{}

%\pagestyle{plain}

%\setcounter{page}{1}

%%%%%%%%%%%%%%%%%%%%%%%%%%%%%%%%%%%%%%%%%%%%%%%%%%%%%%%%%%%%%%%%%%%%%%%%

\section{Introduction.}

%%%%%%%%%%%%%%%%%%%%%%%%%%%%%%%%%%%%%%%%%%%%%%%%%%%%%%%
In this paper we discuss  diffractive production in the deep inelastic 
scattering in the framework of CGC/saturation approach (see Ref.\cite{KOLEB}
 for review). In spite of the fact that the equations for the diffractive
 production in this approach, were  proven   long ago \cite{KOLE}(see
 also Ref.\cite{HWS,HIMST,KLW}) the intensive study, during the past
 two decades, has been concentrated on the simplified model in which
 the diffractive production of quark-antiquark pair and one additional
 gluon has been considered (see Refs.\cite{GBKW,GOLEDD,SATMOD0,KOML,
MUSCH,MASC,MAR,KLMV}).  Such models described the experimental data quite
 well,  giving the impression that we do not need to   search for 
the
 solution of the general equations. Indeed, we found only two attempts
 to solve the equations of Ref.\cite{KOLE} numerically
 (see Refs,\cite{LELUDD,LELUDD1}).

The main goal of this paper is to investigate the non-linear equations for 
 diffractive production in DIS, and to find an analytical solution in 
different
 kinematic regions. Based on these analytical solutions we will
 suggest
 an impact parameter dependent model in the spirit of 
 Refs.\cite{CLP1,CLP} which is
 based on Color Glass Condensate/saturation  effective
 theory  for high energy QCD.

   The paper consists of two parts. In the first part, which is
 the most important  contribution in this paper,  we found the 
analytical solutions
 of the evolution equations for  diffraction production\cite{KOLE}
 in different kinematic regions, mostly using the approach developed in
 Ref.\cite{LETU}.
In the second part of the paper, we suggest an interpolation formula 
which satisfies
 two limits found  analytically: 
deep in the saturation region, and in the vicinity of the
 saturation scale, putting into practice the key ideas of Ref.\cite{IIM}.
This formula exhibits  the main features of the  DIS amplitude ,
 given in Refs.\cite{CLP1,CLP,CLM}, but it is  different from 
 the interpolation procedure that have been used in numerous attempts
 to build a such model  in   Refs. \cite{IIM,SATMOD0,SATMOD1,BKL,
SATMOD2,SATMOD3,SATMOD4,SATMOD5,SATMOD6,SATMOD7,SATMOD8,SATMOD9,SATMOD10,
 SATMOD11,SATMOD12,SATMOD13,SATMOD14,SATMOD15,SATMOD16,SATMOD17}. 
We will attempt to describe  the HERA data in the region
 of  small $x_\pom$ and small $\beta$.\cite{HERADATA}.

Unfortunately,  we are  still doomed to build models to introduce the 
main
 features of the CGC/saturation approach,                                   
since   the CGC/saturation equations do not   reproduce the correct
 behavior of the scattering amplitude at large impact parameter
 (see Ref. \cite{KW,FIIM}).   Real progress in theoretical understanding of
 the confinement of quarks and gluon has not yet  been achieved  and,  as a
 result, we do not know how to  formulate the CGC/saturation 
equations to
 incorporate the phenomenon of  confinement.  We have to build a  model
 which includes both the  theoretical knowledge that stems from the
 CGC/saturation equations,  and  the phenomenological large $b$ behavior
 that does not contradict   theoretical restrictions \cite{FROI,BRLE}. 
 In our modeling of the large $b$ behaviour of the scattering amplitude,
 we follow the main ideas of all saturation models on the market (see for
 example Refs.\cite{IIM,SATMOD0,SATMOD1,BKL,SATMOD2,SATMOD3,SATMOD4,SATMOD5,
SATMOD6,SATMOD7,SATMOD8,SATMOD9,SATMOD10, SATMOD11,SATMOD12,SATMOD13,SATMOD14,
SATMOD15,SATMOD16,SATMOD17}): and only introduce the non-perturbative 
behaviour in the
 $b$-dependence of the saturation scale.

For the $b$ behavior we use the procedure, suggested in Ref.
 \cite{CLP}\footnote{The energy dependence is determined
 theoretically, and in the leading order of perturbative
 QCD $\lambda \,=\, \bas \kappa \,= \,\bas\,\chi\Lb 1 -
 \gamma_{cr}\Rb/(1 - \gamma_{cr}$, where $\chi\Lb \gamma\Rb$
 and $\gamma_{cr}$ are given by \eq{Z}.}:   
 \beq \label{QSB}
  Q^2_s\Lb Y, b\Rb \,=\,Q^2_0 \Lb S\Lb b, m \Rb\Rb^{\frac{1}{\bar \gamma}}
\,e^{\lambda\,Y}
 \eeq
 where     $S\Lb b \Rb $ is the Fourier  image of $ S\Lb Q_T\Rb =
 1/\Lb 1 +    \frac{Q^2_T}{m^2}\Rb^2$,        and we will discuss below 
the 
value of
 $\bar \gamma$. \eq{QSB} leads to the scattering amplitude which  is 
 proportional $\exp\Lb - m b\Rb$ at $b \gg 1/m$ in
 accord with the Froissart theorem \cite{FROI}. In addition,
 we reproduce the large $Q_T$ dependence of this amplitude, which is
 proportional to $Q^{-4}_T$ and follows from the perturbative
 QCD calculation \cite{BRLE}.                                                                                                                                                                                                                                                                                                                                                                                                                                                                                                                                        
  This impact parameter behaviour is the main phenomenological
 assumption that we used.

%%%%%%%%%%%%%%%%%%%%%%%%%%%%%%%%%%%%%%%%%%%%%%%%%%%%%%%%%%%%%%%%%%%%%%%%

\section{Theoretical input}
%%%%%%%%%%%%%%%%%%%%%%%%%%%%%%%%%%%%%%%%%%%%%%%%%%%%%%%%%%%%

In this section we  discuss our theoretical input that follows
 from the Colour Glass Condensate(CGC)/saturation effective theory of QCD
 at high energies(see Ref.\cite{KOLEB} for the basic introduction).

\subsection{ The evolution equation for diffraction production in the
 framework of CGC.}
\begin{figure}[ht]
    \centering
  \leavevmode
      \includegraphics[width=16cm]{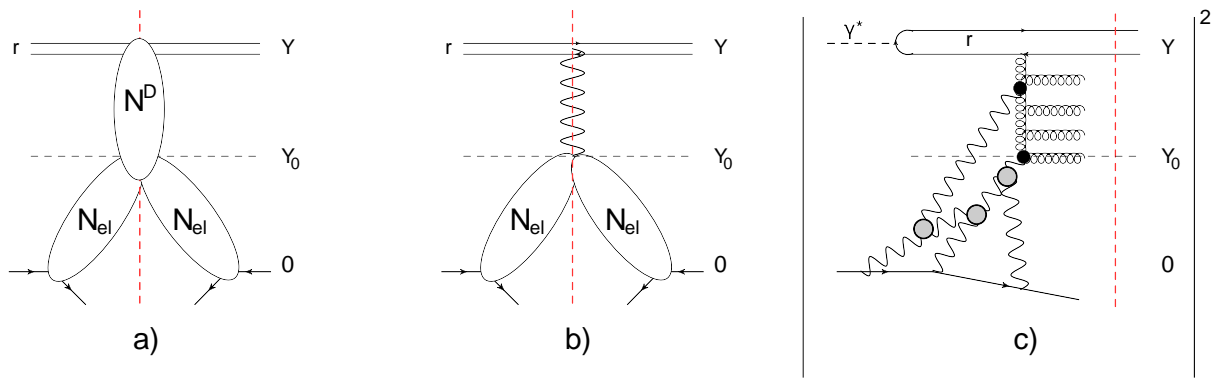}  
      \caption{The graphic representation of the processes of
 diffraction production
}
\label{gen}
   \end{figure}

%%%%%%%%%%%%%%%%%%%%%%%%%%%%%%%%%%%%%%%%%%%%%%%%%%%%%%%%%%%
A sketch of the process of diffraction production in DIS is shown
 in \fig{gen}-c,  from this figure one can see that the main formula 
takes the form
\beq \label{EQ1}
 \sigma^{\rm diff}(Y, Y_0, Q^2)\,\,\,=\,\,\int\,\,d^2
r_{\perp} \int \,d z\,\, |\Psi^{\gamma^*}(Q^2; r_{\perp}, z)|^2
\,\,\sigma_{\rm dipole}^{diff}(r_{\perp}, Y, Y_0)\,, \eeq
where $Y = \ln\Lb 1/x_{Bj}\Rb$ and $Y_0$ is the minimum rapidity gap for
 the diffraction process (see \fig{gen}-c). In other words, we 
consider  diffraction production, in which all produced hadrons have 
rapidities larger than $Y_0$. For $\sigma_{\rm dipole}^{diff}(r_{\perp},
 Y, Y_0)$ we have a general expression
\beq \label{EQ2}
 \sigma_{\rm dipole}^{diff}(r_{\perp},Y, Y_0)
\,\,=\,\,\,\,\int\,d^2 b\,d^2 b'\,N^D(r_{\perp},Y, Y_0;\vec{b})\,, 
\eeq
where the structure of the amplitude $N^D$ is shown in  \fig{gen}-a.

 %%%%%%%%%%%%%%%%%%%%%%%%%%%%%%%%%%%%%%%%%

 %%%%%%%%%%%%%%%%%%%%%%%%%%%%%%%%%%%%%%%%%%%%%%

For $N^D$ the evolution equation has been derived in Ref.\cite{KOLE} in the
 leading log(1/x) approximation (LLA) of perturbative
 QCD(see Ref.\cite{KOLEB} for details and general
 descriptions of the LLA). Hence, we hope
 to describe the experimental data only in the kinematic
 region where both $\beta$ and $x_\pom$ are very small
 ($Y - Y_0 \,=\,\ln(1/\beta) \,\gg\,1$ and $Y_0 = \ln(1/x_{pom}
 \,gg\,1$ are large.  We are aware, that it is sufficient to describe most
of th experimental data by only taking into account
    $q \bar q$ and $q \bar{q} G$ final states in 
 diffraction production.  Because of this, our main goal is not
 to describe the current
 experiments, but to study the solution to the equations in the LLA,
 which introduces the screening corrections to  all the channels of
 the diffraction production.  By comparing with the
 experimental data, we wish to determine in which kinematic region the 
shadowing corrections
 will become important, both for the elastic amplitudes in \fig{gen}, as
 well as for the diffractive  production of the large number of  gluons.

The equation as has been shown in Ref.\cite{KOLE},  can be written in
 two forms. First, it turns out that for the new function
\beq \label{EQ3}
{\cal N}\Lb Y, Y_0; r_{\perp}, b \Rb\,\,\equiv\,\,2 N_{\rm el}\Lb Y;
 r_{\perp}, b \Rb \,\,-\,\,N^D(r_{\perp},x,x_0;b)
\eeq
the equation has the same form as Balitsky-Kovchegov equation\cite{BK}:
 viz.
\bea \label{EQ4}
\displaystyle{\frac{\partial {\cal N}\Lb Y, Y_0; \vec{x}_{01}, \vec{b}
 \Rb}{\partial Y}}\,&=&\,\displaystyle{\frac{\bas}{\pi}\int d^2  {\mathbf{x_{2}}}\frac{{\mathbf{x^2_{01}}}}{{\mathbf{x^2_{02}}}\,
{\mathbf{x^2_{12}}}}\Bigg\{ {\cal N}\Lb Y,Y_0; \vec{x}_{02},\vec{b} - \h \vec{x}_{12}\Rb +  {\cal N}\Lb Y,Y_0; \vec{x}_{12},\vec{b} - \h \vec{x}_{02}\Rb\,-\,{\cal N}\Lb Y,Y_0; \vec{x}_{01},\vec{b} \Rb}\,\nn\\
&-&\,\displaystyle{{\cal N}\Lb Y,Y_0; \vec{x}_{02},\vec{b} - \h \vec{x}_{12}\Rb {\cal N}\Lb Y,Y_0; \vec{x}_{12},\vec{b} - \h \vec{x}_{02}\Rb\Bigg\}}
\eea
Note, that $\vec{r} = \vec{x}_{01}$ and the kernel of the equation
 describe the decay of   a dipole to two dipoles: 
$\vec{x}_{01}\,\to\,\vec{x}_{02}\,+\,\vec{x}_{12}$.
 The initial condition for \eq{EQ4} has the following form:
\beq \label{EQ5}
{\cal N}\Lb Y = Y_0, Y_0; \vec{x}_{01}, \vec{b} \Rb\,\,=\,\,2\,N_{\rm el}\Lb Y = Y_0; \vec{x}_{01}, \vec{b} \Rb\,-\,N^2_{\rm el}\Lb Y = Y_0; \vec{x}_{01}, \vec{b} \Rb
\eeq

Re-writing 
\eq{EQ4} as  the equation for $N^D$ we obtain the second form of
 the set of the equations:

\bea \label{EQ6}
&& \frac{   \partial N^D\Lb Y,Y_0; \vec{x}_{01},\vec{b}\Rb}{ \partial Y} \,=\\
&&\,\frac{\bas}{\pi}
 \int \, d^2 \vec{x}_2
\frac{{\mathbf{x^2_{01}}}}{{\mathbf{x^2_{02}}}\,
{\mathbf{x^2_{12}}}}
\Bigg\{\, N^D \Lb Y,Y_0; \vec{x}_{02},\vec{b} - \h \vec{x}_{12}\Rb +  N^D\Lb Y,Y_0; \vec{x}_{12},\vec{b} - \h \vec{x}_{02}\Rb\,-\,N^D\Lb Y,Y_0; \vec{x}_{01},\vec{b} \Rb\nn\\
&& +\,\,  N^D(Y, Y_0; \vec{x}_{02},\vec{ b}- \h\vec{x}_{12} )  N^D(Y,Y_0; \vec{x}_{12},  \vec{ b}- \h\vec{x}_{02}) 
- 4 \,N^D(Y, Y_0; \vec{x}_{02},\vec{ b}- \h\vec{x}_{12} )  N_{\rm el}(Y; \vec{x}_{12},  \vec{ b}- \h\vec{x}_{02}) \nn\\
&&
+2\, N_{\rm el}(Y; \vec{x}_{02},\vec{ b}- \h\vec{x}_{12} )  N_{\rm el}(Y; \vec{x}_{12},  \vec{ b}- \h\vec{x}_{02})]\,\Bigg\}. \nonumber 
\eea

The initial conditions are
\beq \label{EQ7}
 N^D\Lb Y = Y_0 ,Y_0; \vec{x}_{01}, \vec{b}'\Rb\,\,=\,\,N^2_{\rm el}(Y_0; \vec{x}_{01},  \vec{ b})
  \eeq
 A  general feature, is that the amplitude with fixed rapidity 
gap can be calculate as follows
 \beq \label{EQ8}
 n^D\Lb Y  ,\mbox{ rapidity gap}= Y_0; \vec{x}_{01},\vec{b}\Rb\,=\,- \frac{\partial  N^D\Lb Y  ,Y_0; \vec{x}_{01},\vec{b}\Rb }{\partial Y_0}\,=\,
 \frac{\partial {\cal  N}\Lb Y  ,Y_0; \vec{x}_{01},\vec{b}\Rb }{\partial Y_0}\, 
\eeq

 From \fig{gen} one can see  that $ n^D\Lb Y  ,\mbox{ rapidity gap}= Y_0;
 \vec{x}_{01},\vec{b}\Rb\,\,=\,\,\int d^2 b'  { \nu}^D\Lb Y  ,\mbox{
 rapidity gap}= Y_0; \vec{x}_{01},\vec{b,\vec{b}^{\,'}}\Rb$ where
 $\vec{b}^{\,'}$ is the conjugate variable to the momentum transfer
 in \fig{gen} for the amplitudes $N_{el}$.

 %%%%%%%%%%%%%%%%%%%%%%%%%%%%%%%%%%%%%%%%%%%%%%%%%%%%%%%%%%%%%%%%%%%%%%%%%%%%%%%%%%%%%%%%%%%%%%%%%%%%%%
 \begin{boldmath}
\subsection{$N^D$ deep in the saturation region: $ r^2 Q^2_s\Lb Y
_0 \Rb \,\gg \,1$    and $ r^2 Q^2_s\Lb Y - Y_0\Rb \,\gg \,1$, 
  and a violation of the geometric scaling behavior}
\end{boldmath}
%%%%%%%%%%%%%%%%%%%%%%%%%%%%%%%%%%%%%%%%%%%%%%%%%%%%%%%%%%%%%%%%%%%%%%%%%%%%%%

First, we consider the kinematic region, where 
       $ r^2 Q^2_s\Lb Y_0 \Rb \,\gg \,1$   
 and $ r^2 Q^2_s\Lb Y - Y_0\Rb \,\gg \,1$.
  Note, that  $ r^2 Q^2_s\Lb Y \Rb\, \gg\, 1$ stems from the above
 restrictions.  
     
 In this  region where both $Y$ and $Y_0$  as well as the
 difference between them are large, we can expect that both $N^D$ and 
$N_{\rm el}$
 are close to 1. Therefore, we can use the procedure suggested in Ref.
     \cite{LETU}. In this region we can replace 
     \beq \label{DSR1}
     {\cal N}^D\Lb Y,Y_0, \vec{x}_{01},\vec{b}\Rb\,\,=\,\,
1 \,-\,\Delta^D\Lb Y,Y_0, \vec{x}_{01},\vec{b}\Rb ;~~~~~~~~~~~~   
N_{\rm el}\Lb Y, \vec{x}_{01},\vec{b}\Rb\,\,=\,\,1 \,-\,\Delta_{\rm el}\Lb Y, \vec{x}_{01},\vec{b}\Rb ;     
     \eeq
     and linearize  \eq{EQ4}, neglecting $\Lb \Delta^D\Rb^2$ terms.
  Indeed, \eq{EQ4} takes the form
     
     \beq \label{DSR2}
\displaystyle{\frac{\pp \Delta^{D}_{01}}{\pp Y} 
}=\displaystyle{\frac{\bas}{\pi}\int d^  
  {2}x_{2}\frac{x^{2}_{01}}{x^{2}_{02}x^{2}_{21}}[ 
\Delta^{D}_{02}\Delta^{D}_{12}-\Delta^{D}_{10} ]   } \eeq     
where we use  notation $\Delta^D_{ik} \equiv \Delta^D\Lb Y,Y_0,
 \vec{x}_{ik},\vec{b}\Rb$ and considered in \eq{DSR2} the impact
 parameter $|\vec{b}| \,\gg\,|\vec{x}_{02}|$ and
 $|\vec{x}_{12}|$. The initial condition  to \eq{DSR2}, given by \eq{EQ5},
 can be re-written in the form
 \beq \label{DSR3}
 \Delta^{D}(Y=Y_{0},\vec{x}_{10},\vec{b})\,\,=\,\,\mbox{C}^{2}\,\displaystyle{\exp\left\lbrace-\frac{\ln^2 \Lb x^2_{10} \,Q^2_s\Lb Y_0; b\Rb \Rb}{\kappa}   \right\rbrace}  
\eeq
In \eq{DSR3} we use the solution given 
in Re.\cite{LETU} for $\Delta_{\rm el}$ which has the form

\beq \label{DSR4}
 \Delta_{\rm el}(Y; \vec{x}_{10},\vec{b})\,\,=\,\,\mbox{C}\,\displaystyle{\exp\left\lbrace-\frac{\ln^2 \Lb x^2_{10} \,Q^2_s\Lb Y; b \Rb \Rb}{2\,\kappa} \right\rbrace}   \eeq
 
 In \eq{DSR3} and \eq{DSR4} $Q_s\Lb Y,b\Rb$ is the saturation scale. 
 Both these equations show the geometric scaling behaviour of the
    scattering amplitude \cite{GS,IIML,LETU}, which depends on a single 
variable
 
\bea \label{Z}
z\,\,&=&\,\,\ln\Lb x^2_{10}\,Q^2_s\Lb Y,b \Rb\Rb\,\,=\,\,\bas\,\kappa \,Y \,+\,\ln\Lb Q^2_s\Lb Y=0,b\Rb\,x^2_{10}\Rb;\nn\\
z_0\,&=&\,\ln\Lb x^2_{10}\,Q^2_s\Lb Y_0, b \Rb\Rb\,\,=\,\,\bas\,\kappa \,Y_0 \,+\,\ln\Lb Q^2_s\Lb Y=0,b\Rb\,x^2_{10}\Rb;\nn\\
\kappa &=&\,\,\frac{\chi\Lb 1 - \gamma_{cr}\Rb}{1 - \gamma_{cr}}; \,\, \,\, \chi\Lb \gamma\Rb \,=\,2 \psi\Lb 1\Rb - \psi\Lb \gamma\Rb - \psi\Lb 1 - \gamma\Rb;
\eea
 where $\psi(x) = d \ln \Gamma(x)/d x$  and $\Gamma$ is
 the Euler gamma function \cite{RY}. In this paper we will use
 the value of $\gamma_{cr} $ which comes from the leading order estimates: 
   $ \gamma_{cr}\,\approx\,\,0.37$.

    Neglecting the term proportional to $(\Delta^D)^2$ in \eq{EQ4} and
 integrating over $\vec{x}_2 $  from $1/Q_s\Lb Y,b\Rb$ to $\vec{x}^2_{10}$.
  The linear equation can be multiplied by arbitrary function of $Y_0$ and
 $x_{10}$. Bearing this in mind, the solution has the following form:
    \beq \label{DSR5}
     \displaystyle{ \Delta^{D}( Y, Y_0; x_{10} ) =G(Y_{0}, x^2_{10})\,\,e^{-z^{2}/2\kappa}   }
     \eeq 
     
    Note that function $G$ can be found from the initial condition of
 \eq{DSR3}, leading to the final answer:
     \beq \label{DSR6}
      \Delta^D\Lb Y,Y_0; x_{10}\Rb \,\,=\,\,  \mbox{C}^{2}\exp\Lb -\frac{z^2_0}{2 \,\kappa} - \frac{z^2}{2 \kappa} \Rb  ;
      \eeq 
      
      The most impressive feature of the solution is that the function
 does not show the geometric scaling behaviour i.e. being a function of 
one
 variable. The solution is the product of two functions: one has a
 geometric scaling behaviour depending on one variable $z$, and the
 second depends on $z_0$, showing  geometric scaling behaviour in
 the same way, as  elastic scattering amplitude at $Y=Y_0$.
        
%%%%%%%%%%%%%%%%%%%%%%%%%%%%%%%%%%%%%%%%%%%%%%%%%%%%%%%%%%%%

%%%%%%%%%%%%%%%%%%%%%%%%%%%%%%%%%%%%%%%%%%%%%%%%%%%%%%%%%%%%%%%%%%%%%%%%%%%%%%%%%%%%%%%%%%%%%%%
     \begin{figure}[ht]
    \centering
  \leavevmode
      \includegraphics[width=9cm]{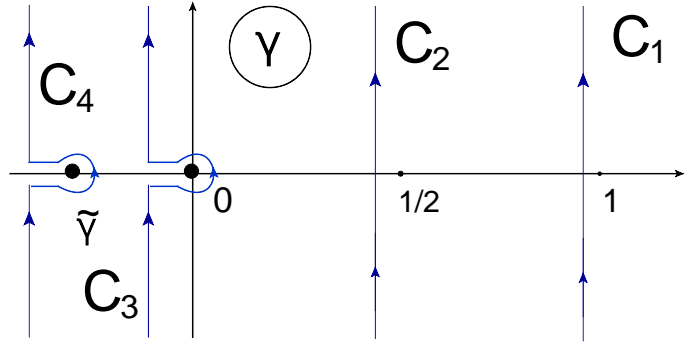}  
      \caption{The contours of integration over $\gamma$}
\label{cntr}
   \end{figure}

 %%%%%%%%%%%%%%%%%%%%%%%%%%%%%%%%%%%%%%%%%%%%%%  %%%%%%%%%%%%%%%%%%%%%%%%%%%%%%%%%%%%%%%%%%%%%%%%% 

  ~
      
      ~
     
\begin{boldmath}
\subsection{ Vicinity of the saturation scale at $r^2\,Q^2_s\Lb Y_0, b\Rb\, \approx \,1$}
\end{boldmath}

%%%%%%%%%%%%%%%%%%%%%%%%%%%%%%%%%%%%%%%%%%%%%%%%%%%%%%%%%%%%
In this subsection we consider the kinematic region in which
 $N_{\rm el}\Lb Y_0, \vec{x}_{01}, \vec{b}\Rb$ is in the vicitity of
 the saturation scale $Q_s\Lb Y_0,b \Rb$, but at $x^2_{01}\,Q^2_s\Lb 
Y_0,b\Rb \, <\, 1$. As it was found in Ref. \cite{IIML} we have  geometric
 scaling behaviour in this region, and the amplitude behaves as
\beq \label{V1}
\displaystyle {N_{\rm el}\Lb Y_0, \vec{x}_{01}, \vec{b}\Rb  \,\,\propto\,\,\Lb x^2_{10}\,Q^2_s\Lb Y_0,b\Rb\Rb^{1 - \gamma_{cr}} }
\eeq

If $Y $ is close to $Y_0$, we can neglect the non-linear
 terms in \eq{EQ6}, and we can solve the linear equation for $N^D$
 with the initial condition of \eq{EQ7}
\beq \label{V2}
\displaystyle {N^D\Lb Y_0, Y_0; \vec{x}_{01}, \vec{b}\Rb  \,\,=\,\,\mbox{c}^2\,\Lb x^2_{10}\,Q^2_s\Lb Y_0,b\Rb\Rb^{2(1 - \gamma_{cr})} }    ;
\eeq   

%%%%%%%%%%%%%%%%%%%%%%%%%%%%%%%%%%%%%%%%%%%%%%%%%%%%%%

~

\begin{boldmath}
\subsubsection{Solution in the region where $N_{\rm el} \,\,<\,\,1$}
\end{boldmath}
%%%%%%%%%%%%%%%%%%%%%%%%%%%%%%%%%%%%%%%%%%%%%%%%%%%%%%
Considering  \eq{V2}, one can see that in this kinematic region we can 
 in general neglect the  expression of \eq{EQ6} two terms:  the term which 
is
 proportional to $\Lb N^D\Rb^2$  at $Y-Y_0 \ll Y_0$, since it  is
 of the order of $N^4_{\rm el}$,  and the term which is proportional
 to $N^D N_{\rm el} \propto N^3_{\rm el}$,  while we have to keep all
 other terms.

 Therefore, the equation takes the form:

\bea \label{V3}
&& \frac{   \partial N^D\Lb Y,Y_0; \vec{x}_{01},\vec{b}\Rb}{ \partial Y} \,=\\
&&\,\frac{\bas}{\pi}
 \int \, d^2 \vec{x}_2
\frac{{\mathbf{x^2_{01}}}}{{\mathbf{x^2_{02}}}\,
{\mathbf{x^2_{12}}}}
\Bigg\{\, N^D \Lb Y,Y_0; \vec{x}_{02},\vec{b} - \h \vec{x}_{12}\Rb +  N^D\Lb Y,Y_0; \vec{x}_{12},\vec{b} - \h \vec{x}_{02}\Rb\,-\,N^D\Lb Y,Y_0; \vec{x}_{01},\vec{b} \Rb\nn\\
&&
+2\, N_{\rm el}(Y; \vec{x}_{02},\vec{ b}- \h\vec{x}_{12} )  N_{\rm el}(Y; \vec{x}_{12},  \vec{ b}- \h\vec{x}_{02})]\,\Bigg\}. \nonumber 
\eea

In this equation we take into account the corrections of the order
 $N^2_{el}$, but neglected  the terms of the order of $N^3_{el}$ and
 $N^4_{el}$, assuming they are small. We believe that this equation will 
allow us to take into account the correction for $N_{el} \approx 0.4 - 
0.5$,

Taking derivatives  with respect to $Y_0$, we re-write \eq{V3} for the
 amplitude $n^D\Lb Y,Y_0,\vec{x}_{01},\vec{b}\Rb$ that has been introduced
 in \eq{EQ8}. It takes the form of the linear equation:
\bea \label{V4}
&& \frac{   \partial n^D\Lb Y,Y_0; \vec{x}_{01},\vec{b}\Rb}{ \partial Y} \,=\\
&&\,\frac{\bas}{\pi}
 \int \, d^2 \vec{x}_2
\frac{{\mathbf{x^2_{01}}}}{{\mathbf{x^2_{02}}}\,
{\mathbf{x^2_{12}}}}
\Bigg\{\, n^D \Lb Y,Y_0; \vec{x}_{02},\vec{b} - \h \vec{x}_{12}\Rb +  n^D\Lb Y,Y_0; \vec{x}_{12},\vec{b} - \h \vec{x}_{02}\Rb\,-\,n^D\Lb Y,Y_0; \vec{x}_{01},\vec{b} \Rb\Bigg\} \nn
\eea
The initial condition for this equation is the following:
\beq \label{V5}
n^D\Lb Y = Y_0,Y_0; \vec{x}_{01},\vec{b}\Rb\,\,=\,\,\frac{\partial
}{\partial Y_0} N^2_{el}\Lb Y_0, \vec{x}_{01},\vec{b}\Rb
\eeq

The elastic amplitude has the form:
  \beq \label{NELA}
N_{\rm el}(Y; x_{10},  b) \,\,=\,\,{\rm c}\Lb x^2_{10}\,Q^2_s\Lb Y,b\Rb\Rb^{\bar{\gamma}}\,\,\equiv\,\,{\rm c}\,\Lb \frac{Q^2_s\Lb Y_0,b\Rb }{Q^2_0}\Rb^{\bar{\gamma}}\,e^{\bar{\gamma} \Lb \bas \kappa \Lb Y - Y_0\Rb \,-\, \xi\Rb}\eeq  
where $\xi\,\equiv\,\ln\Lb 1/\Lb x^2_{10}\,Q^2_0\Rb\Rb$.

  Taking the double Mellin transform, and defining $\bar{Y}=\bas Y$
  \beq \label{DM}
\displaystyle{n^D\Lb Y, Y_0, \xi, b\Rb\,\,=\,\,\, \int_{C_1}\frac{
 d \gamma d \omega}{ (2 \pi i)^2}\,\phi(\omega, \gamma)\,e^{ \omega
 \Lb \bar{Y}- \bar{Y}_0\Rb\,\,+\,\,\Lb \gamma - 1\Rb \xi}}
\eeq  
we obtain the solution to the  equation of \eq{V4} in the following form:
 \beq \label{DMS}
\displaystyle{n^D\Lb Y, Y_0, \xi, b\Rb\,\,=\,\,\, \int_{C_1}\frac{ d \gamma d \omega}{ (2 \pi i)^2}\,\frac{\phi_{in}\Lb \gamma,Y_0,\vec{b}'\Rb}{\omega - \chi\Lb \gamma\Rb}\,e^{ \omega \Lb \bar{Y}- \bar{Y}_0\Rb\,\,+\,\,\Lb \gamma - 1\Rb \xi}}
\eeq  
where $\phi_{in}$ has to be determined from the initial condition of
 \eq{V5}, and it has the form
\beq \label{DM1}
\phi_{in}\Lb \gamma,Y_0,\vec{b}'\Rb\,\,=\,\,2\bar{\gamma}\kappa\,\Big(\frac{Q^2_s\Lb Y_0, b'\Rb}{Q^2_0}\Big)^{2 \bar{\gamma}}\,\frac{c^2}{\gamma - \tilde{\gamma}}
\eeq
Therefore, the solution takes the  form  (see \fig{gen} for  notations):

\beq \label{CR1}
\displaystyle{n^D\Lb Y, Y_0, \xi, b\Rb\,\,=\,\,2\bas\,\bar{\gamma}\,\kappa\,c^2\Big(\frac{Q^2_s\Lb Y_0, b'\Rb}{Q^2_0}\Big)^{2 \bar{\gamma}}
\int_{C_1}\frac{ d \gamma}{ 2 \pi i}\,
\frac{1}{ \gamma - \tilde{\gamma}}\,e^{ \chi\Lb \gamma\Rb \Lb  \bar{ Y} - \bar{Y}_0\Rb\,\,+\,\,\Lb \gamma - 1\Rb \xi}}\eeq
with  $\gamma_{cr} = 0.37$ and $\bar{\gamma}\,=\,\,1 - \gamma_{cr}=0.63$ ,$ \tilde{\gamma} = -1+ 2 \gamma_{cr} =  -  0.26$.  $\chi\Lb \gamma\Rb $ is given by \eq{Z}.

The choice of the contour of integration over $\gamma$ (see \fig{cntr})
is standard for the solution of the BFKL Pomeron, and correctly reproduces
  the calculation of the gluon emission in perturbative QCD.

The contour of integration ($C_1$) is shown in \fig{cntr}. Since $
\xi > 0$ we can safely move this contour, and for large values of
 $\delta Y = Y - Y_0$ and $\xi$,  we can take the integral using
 the method of steepest decent. For $\as \delta Y \,\gg\,\xi$ we
 evaluate the integral by this method,   integrating along  the contour
 $C_2$ which crosses the real axis at  $\gamma$ close to $\h$.
 At $  \xi\,\gg \,\bas\,\delta Y  $, we can integrate by the same
 method, but moving contour $C_2$  closer to $y$-axis in \fig{cntr}.
 For $\delta Y = 0$ we can close the counter over pole
 $\gamma = \bar{\gamma}$. However, for   $\delta Y \sim 1$ we
 cannot use the same method, since at $\gamma = 0$ we have singularities 
in
 the kernel $\chi(\gamma)$.
 We cannot use the method of 
steepest decent for such small values of $\delta Y$.

 %%%%%%%%%%%%%%%%%%%%%%%%%%%%%%%%%%%%%%%%%%%%%%%%%%%%%%
~
\subsubsection{Solution in the saddle point approximation}

%%%%%%%%%%%%%%%%%%%%%%%%%%%%%%%%%%%%%%%%%%%%%%%%%%%%%% 

Using solution of \eq{CR1} we can find the saturation momentum for
 the process of  diffraction production, calculating the integral
 over $\gamma$  by the method of steepest descent.  $Q_s$  can be
 determined from the following two equations:
\bea
\bas \frac{ d \chi\Lb \gamma\Rb}{d \gamma} \Lb Y - Y_0\Rb\,\, + \,\,\xi &=& \,\,0;      \label{QS11}\\
\bas\chi\Lb \gamma\Rb \Lb Y - Y_0\Rb \,\,+\,\, \Lb \gamma - 1\Rb \xi &=& \,\,0;\label{QS12}
\eea
  \eq{QS11} is the equation for the saddle point  while \eq{QS12} is the condition that the solution is a constant on the critical line $x^2_{10} = 1/Q_s^2$. The solution of these two equation is well known  $\gamma = \gamma_{cr} = 0.37$ and $Q^2_s\Lb Y\Rb \, =\, Q^2\Lb Y_0\Rb\,\exp\Lb \bas \kappa \Lb Y - Y_0\Rb\Rb$     with $\kappa=\chi\Lb \gamma_{cr}\Rb/(1 - \gamma_{cr})$. In the vicinity of the saturation scale  \eq{CR1} behaves as
  \beq \label{QS3}
  \frac{2\,\bas\, \bar{\gamma}\,\kappa\,c^2}{\gamma_{cr} - \tilde{\gamma}}\,\Lb \frac{Q^2\Lb Y_0,b'\Rb }{Q^2_0}\Rb^{2\,\bar{\gamma}}\Lb x^2_{10}\,Q^2_s\Lb Y - Y_0, \vec{b} - \vec{b}'\Rb\Rb^{1 - \gamma_{cr}}
 \eeq 
 
One can see that there is no geometric scaling behavior
 of the scattering amplitude $n^D$, even at large $Y - Y_0$.
 We can also see that the solution does not satisfy the initial condition.
 It stems from \eq{QS11} and \eq{QS12}, which both are correct only if
 $\delta Y \,\gg\,1$, assuming that the saddle point value of $\gamma$
 is not close to $\tilde{\gamma}$.  We need to re-write these equations
 to take into account the possibility that $\gamma_{SP} \to \tilde{\gamma}$
 taking into account the factor $1/(\gamma - \tilde{\gamma})$. The equations
 for the saddle point  take the form:

\bea
\bas \frac{ d \chi\Lb \gamma\Rb}{d \gamma} \Lb Y - Y_0\Rb\,\, + \,\,\xi \,\,+\,\,\frac{1}{\gamma - \tilde{\gamma}}&=& \,\,0;      \label{QS21}\\
\bas\chi\Lb \gamma\Rb \Lb Y - Y_0\Rb \,\,+\,\, \Lb \gamma - 1\Rb \xi \,-\,\ln\Lb \gamma - \tilde{\gamma}\Rb&=& \,\,0;\label{QS22}
\eea

The contribution of the additional term is essential, only if
 $\gamma_{SP} \to \tilde{\gamma}$, but even $\gamma=0$ which
 corresponds the integration with the contour $C_3$ (see \fig{cntr}),
 is still not close to $\tilde{\gamma}$.  Bearing this in mind, we prefer
 to treat $\delta Y \ll 1$  without using the method of steepest descent.

~
%%%%%%%%%%%%%%%%%%%%%%%%%%%%%%%%%%%%%%%%%%%%%%%%%%%%%%

\subsubsection{Solution for $\delta Y \ll 1$. }

%%%%%%%%%%%%%%%%%%%%%%%%%%%%%%%%%%%%%%%%%%%%%%%%%%%%%% 

 In this section we will return to the discussion
 of \eq{CR1} at $\delta Y = Y - Y_0\,\leq\,1$.  In this
 kinematic region,  we cannot use the method of steepest
 descent, and have to look for a different  approach. First,
 let us analyze the solution iterating the equation keeping $\delta Y \,\,\ll\,1$. To obtain the solution as a sum of $ \Lb \delta Y\Rb^n$ contributions we need to expand
   \beq \label{V43}
 e^{ \bas \chi\Lb \gamma\Rb \Lb Y - Y_0\Rb}\,\,=\,\,\sum^\infty_{n=0} \,\frac{\Lb \bas \chi(\gamma)\,\delta Y\Rb^n}{n!} \,\xrightarrow{ \gamma \to 0}\,\sum^\infty_{n=0} \,\frac{1}{n!} \Lb \frac{ \bas \,\delta Y}{\gamma}\Rb^n
 \eeq 
 For each term of this series,  we need to plug  in our solution
  and integrate it over $\gamma$. This integral  takes the
 following  form for the third term in \eq{CR1} for $n\,\geq\,1$:
 
 \beq \label{V44}
\oint_{C_3} \,d \gamma \,\Lb \frac{ \bas \,\delta Y}{\gamma}\Rb^n\frac{e^{ \Lb \gamma - 1\Rb \xi} }{ \gamma - \tilde{\gamma}}\,\,=\,\,
\Lb \bas \,\delta Y\Rb^n\Bigg\{\frac{1}{\tilde{\gamma}}\frac{1}{(n - 1)!} \Lb \frac{e^{ \Lb \gamma - 1\Rb \xi} }{ \gamma - \tilde{\gamma}}\Rb^{(n)}_{\gamma,\gamma \to 0}\,+\,\frac{1}{\tilde{\gamma}^n}\,e^{ \Lb\tilde{ \gamma} - 1\Rb \xi}\Bigg\}
\eeq  

For $n=0$, we have the contribution only of the second term in \eq{V44}.

In \eq{V43} we evaluated the integral, closing the contour over
 the singularities of the BFKL kernel, which is the pole at
 $\gamma=0$,  and over the pole $\gamma = \tilde{\gamma}$.
 The BFKL kernel also has poles  at $\gamma\,=\,- n,\, n = 1,2,3 \dots$, 
   but their contributions are exponentially suppressed with $\xi$ leading 
to
 the next twists contributions. \eq{V44} can be re-written as follows
\beq \label{V45}
\oint_{C_3} \,d \gamma \,\Lb \frac{ \bas \,\delta Y}{\gamma}\Rb^n\frac{e^{
 \Lb \gamma - 1\Rb \xi} }{ \gamma - \tilde{\gamma}}\,\,=\,\,
\Lb \frac{\bas \,\delta Y}{\tilde{\gamma}}\Rb^n\, e^{- \xi}\,\Bigg\{\frac{1}{\tilde{\gamma}}\frac{1}{(n - 1)!}\,\sum_{k=0}^{n - 1}\frac{\Lb \tilde{\gamma} \, \xi\Rb^k}{k!} \,\, +\,\,e^{\tilde{\gamma} \,\xi}\Bigg\}
\eeq
 
 The last term in \eq{V45} is the contribution at the pole $\gamma =
 \tilde{\gamma}$, while the first term is the sum of logs term giving
 the leading twist perturbative series.
 
 Bearing this in mind we can re-write \eq{CR1} in the following form
 \beq \label{CR2}
n^D\Lb Y, Y_0,\xi , b\Rb\,\,=\\
\,\,2\bas \bar{\gamma} \,\kappa\,\,c^2\,\Lb \frac{Q^2\Lb Y_0,b'\Rb }{Q^2_0}\Rb^{2\bar{\gamma}}\Bigg\{\int_{C_1 - C_4}\frac{d \gamma}{ 2 \pi i}\,
\frac{1}{ \gamma - \tilde{\gamma}}\,e^{ \chi\Lb \gamma\Rb \Lb  \bar{ Y} - \bar{Y}_0\Rb\,\,+\,\,\Lb \gamma - 1\Rb \xi}
\,\,+\,\,e^{ \Lb\tilde{ \gamma} - 1\Rb \xi}  \,e^{\bas \chi\Lb \tilde{\gamma}\Rb\,
\Lb Y - Y_0\Rb}\Bigg\}
\eeq
  The first term in \eq{CR2} is the difference between two integrals
 with contour $C_1$ and $C_4$, while the last term is the result of
 integrating over $\gamma$, with the contour $C_4$. The advantage
 of this form for the equation, is that it satisfies the initial
 condition, since the first term is equal to zero at $\delta Y =
 0$, and the first term generates all perturbative logs   with
 respect to the dipole sizes.
 
 In the situation when $\bas \xi \,\gg\,1$    while $\bas \,\ll\,1$ 
 the first term reduces to the double log approximation generating
 the contribution
 {\small
\bea \label{V5}
&& n^D_{\rm 1-st\, term \,of\, \eq{CR2}}\Lb Y,Y_0,   r\Rb\,\,=\\
&& 2\bas \bar{\gamma}\,\kappa\, c^2\,\Lb \frac{Q^2\Lb Y_0,b'\Rb }{Q^2_0}\Rb^{2\bar{\gamma}}\frac{\bas \delta Y}{\bar{\gamma}}\,e^{- \xi}\,\sum^\infty_{n=1}\frac{1}{n! (n - 1)!} \Lb \bas \delta Y \xi\Rb^{n - 1}\,
 =\,2 c^2 \,\Lb \frac{Q^2\Lb Y_0,b'\Rb }{Q^2_0}\Rb^{2\bar{\gamma}}\frac{1}{\bar{\gamma}} \h \sqrt{\frac{\bas \delta Y}{\xi}}\,e^{- \xi} I_1\Lb 2 \sqrt{\bas \delta Y\,\xi}\Rb \,\nn  \eea}
 which stems from the term with $\xi^{n - 1}$ in \eq{V45}.
 
 Finally the double log contribution takes the form:
 \bea \label{V6}
 && n^D\Lb Y,Y_0,   r\Rb\,\, =\\
 &&\,2 \bas\, \bar{\gamma}\,\kappa\,c^2 \,\Lb \frac{Q^2\Lb Y_0,b\Rb }{Q^2_s\Lb Y_{\rm in}\Rb}\Rb^{2\bar{\gamma}}\frac{1}{\bar{\gamma}}\Bigg\{ \h \sqrt{\frac{\bas \delta Y}{\xi}}\,e^{- \xi} I_1\Lb 2 \sqrt{\bas \delta Y\,\xi}\Rb \, +\,\,e^{ \Lb\tilde{ \gamma} - 1\Rb \xi}  \,e^{\bas \chi\Lb \tilde{\gamma}\Rb\,
\Lb Y - Y_0\Rb} \Bigg\} \nn
\eea

In appendix A we remove the  assumption that $\bas \xi \gg \,1$.
%%%%%%%%%%%%%%%%%%%%%%%%%%%%%%%%%%%%%%%%%%%%%%%%%%%%%%
\begin{boldmath}
\subsubsection{Solution in the region where $N_{\rm el} \,\sim\,1$}
\end{boldmath}
%%%%%%%%%%%%%%%%%%%%%%%%%%%%%%%%%%%%%%%%%%%%%%%%%%%%%% 
In this kinematic region we need to keep the term which is
 proportional to $N_{el} \,N^D$ in \eq{EQ5} and solve the 
equation which takes the form

\bea \label{V31}
&& \frac{   \partial n^D\Lb Y,Y_0; \vec{x}_{01},\vec{b}\Rb}{ \partial Y} \,=\\
&&\,\frac{\bas}{\pi}
 \int \, d^2 \vec{x}_2
\frac{{\mathbf{x^2_{01}}}}{{\mathbf{x^2_{02}}}\,
{\mathbf{x^2_{12}}}}
\Bigg\{\, n^D \Lb Y,Y_0; \vec{x}_{02},\vec{b} - \h \vec{x}_{12}\Rb +  n^D\Lb Y,Y_0; \vec{x}_{12},\vec{b} - \h \vec{x}_{02}\Rb\,-\,n^D\Lb Y,Y_0; \vec{x}_{01},\vec{b} \Rb\nn\\
&&-\,4 \,n^D(Y, Y_0; \vec{x}_{02},\vec{ b}- \h\vec{x}_{12} ) \,\, N_{\rm el}(Y; \vec{x}_{12},  \vec{ b}- \h\vec{x}_{02}) \,
\Bigg\}. \nonumber 
\eea
Therefore, we took into account terms $n^D\,N_{el}$ in comparison
 with the previous sections. We consider that $n^D\,N^D$ are 
sufficiently small, so
  that we can neglect the contributions $n^D \,N^D\,\sim N^4_{el} \,\ll 1$.

This equation looks simpler in the 
 momentum representation:
 \beq \label{V41}
N\Lb x_{10},b, Y\Rb\,\,\equiv\,\,x^2_{10}\int d^2 \,k_T e^{i \vec{k}_T \cdot \vec{x}_{10}}N\Lb k_T, b, Y\Rb
\eeq 
 
 In the momentum representation 
\eq{EQ6} takes the form:
\bea \label{V51}
&& \frac{   \partial n^D\Lb Y,Y_0; \vec{k}_T,\vec{b}\Rb}{ \partial Y} \,=\\
&&\,\frac{\bas}{\pi}
 \int \, d^2 \vec{k}'_T\frac{1}{\Lb \vec{k}'_T - \vec{k}_T\Rb^2}\Bigg\{\, n^D \Lb Y,Y_0; \vec{k'}_T,\vec{b} \Rb -\,\frac{k^2_T}{\Lb \vec{k}'_T - \vec{k}_T\Rb^2 + k'^2_T}\,n^D\Lb Y,Y_0; \vec{k}_{T},\vec{b} \Rb \Bigg\}\,   \nonumber   \\
 &&\,+\,\bas\,2 n^D\Lb Y,Y_0; \vec{k}_{T},\vec{b} \Rb N^D\Lb Y,Y_0; \vec{k}_{T},\vec{b} \Rb   \,\,
-\bas\, \,4 \,n^D(Y, Y_0; \vec{k}_T,\vec{ b} )  N_{\rm el}(Y; \vec{k}_T,  \vec{ b}) 
 \,\nonumber 
\eea

   We  solve this equation using the semi-classical approach. In this
 approach we are looking for the solution in the form
   \beq \label{SC1}
  n^D\Lb Y,Y_0; k_T,b\Rb\,\,\equiv\,\, n^D\Lb Y,Y_0; \rho = \ln\Lb k^2_T/Q^2_0\Rb,b\Rb  \,=\,e^{S\Lb Y,Y_0,\rho,b\Rb}
  \eeq
  with
 \beq \label{SC2}
 \,\,\,S\,=\,\omega\Lb Y,Y_0,\rho,b\Rb \Lb Y - Y_0\Rb\,-\,   \Lb 1 - \gamma \Lb Y,Y_0,\rho,b\Rb  \Rb\,\rho
   \eeq
   where 
   
   \beq \label{OMGA}
   \frac{\partial S\Lb Y,Y_0,\rho,b\Rb}{\partial Y}   \,\,=\,\, \omega\Lb Y,Y_0,\rho,b\Rb;
   ~~~~~~~~~~~~~\frac{\partial S\Lb Y,Y_0,\rho,b\Rb}{\partial \rho }\,\,=\,\,\gamma\Lb Y,Y_0,\rho,b\Rb -1;
   \eeq
   are smooth functions of
$Y$ and $\rho$, and the following conditions are assumed:
\bea
&& \frac{\D \, \omega(Y,Y_0,\rho,b)}{\D Y}\hspace{0.3cm}\ll\hspace{0.3cm}\omega^2(Y,Y_0,\rho,b);\,\,\,\,\frac{\D \, \omega(Y,Y_0,\rho,b)}{\D \rho}
\hspace{0.3cm}\ll\hspace{0.3cm}\omega(Y,Y_0,\rho,b)\,\Lb 1 - \gamma(Y,Y_0,\rho,b \Rb;\,\,\, \label{cond2}\\
\nn\\
&&\frac{\D\,
\gamma(Y,Y_0,\rho,b)}{\D \rho} \hspace{0.3cm}\ll\hspace{0.3cm}\Lb 1 - \gamma(Y,Y_0,\rho,b)\Rb^2;\,\,\,\,\,\frac{\D \,
\gamma(Y,Y_0,\rho,b)}{\D Y }
\hspace{0.3cm}\ll\hspace{0.3cm}\omega(Y,Y_0,\rho,b)\,\Lb 1 - \gamma(Y,Y_0,\rho,b)\Rb;\label{cond3}\eea   

Plugging \eq{SC1} and \eq{SC2} in \eq{V5} we have
\beq \label{SC3}
\omega\Lb Y,Y_0,\rho,b\Rb\,\,=\,\,\bas \,\Lb \chi\Lb \gamma\Lb Y,Y_0,\rho,b\Rb\Rb\,\,-\,\,4 N_{\rm el}(Y; \rho,  b)\Rb\,\,
\eeq   
For the equation in the form
\beq \label{SC4}
F(Y,\rho,S,\gamma,\omega)=0
\eeq
where $S$ is given by \eq{SC2}, we can introduce the set
of
characteristic lines on which  $\rho(t), Y(t), S(t),$ $ \omega(t),$ and $
\gamma(t)$ are
functions of the variable $t$ (which we call artificial time),
 that satisfy the following
equations:
\begin{eqnarray}
&&(1.)\,\,\,\,\,\,\,\frac{d \rho}{d\,t}\,\,=\,\,F_{\gamma}\,\,=\,-\,\bas\,\frac{d \chi(\ga)}{d \ga}\nn\\
\,\,\,\,\,\,\,\,\,\,\,\,&&(2.)\,\,\,\,\,\,\,
\frac{d\,Y}{d\,t}\,\,=\,\,F_{\omega}\,\,=\,\,1\,\,\,\,\,\,\,\nn\\
 &&(3.)\,\,\,\,\,\,\,
\frac{d\,\tilde{S}}{d\,t}\,\,=\,\,(\gamma\,-\,1)\,F_{\gamma}\,+\,\omega\,F_{\omega}\,\,=\,\,  \bas\,\Lb 1 \,-\,\ga\Rb\,\frac{d \chi(\ga)}{d \ga}\,\,+\,\,\om\nonumber \\
&&(4.)\,\,\,\,\,\,\,\frac{d\,\gamma}{d\,t}\,\,=\,\,-
(\,F_{\rho}\,+\,(\gamma\,-\,1)\,F_{S}\,)\,\,=\,\, 4\bas (1 - \gamma_{cr})\,N_{\rm el}(Y; \rho,  b) \nn\\
&& (5.)\,\,\,\,\,\,\,
\frac{d\,\omega}{d\,t}\,\,=\,\,- \,(\,F_{Y}  \,+\,\omega\,F_S\,)\,\,= -\,\,4 \bas^2 \kappa \bar{\gamma}\,N_{\rm el}(Y; \rho,  b) \,\label{SCEQ}
\end{eqnarray}

In \eq{SCEQ} we consider that 
\beq \label{NEL}
N^2_{\rm el}(Y; \rho,  b) \,\,=\,\,{\rm c}\Lb \frac{Q^2_s\Lb Y,b\Rb}{k^2_T}\Rb^{\bar{\gamma}}\,\,\equiv\,\,{\rm c}\,\Lb \frac{Q^2\Lb Y_0,b\Rb }{Q^2_0}\Rb^{\bar{\gamma}}\,e^{\bar{\gamma} \Lb \bas \kappa \Lb Y - Y_0\Rb \,-\, \rho\Rb}
\eeq
with $\kappa= \chi\Lb \gamma_{cr}\Rb/(1 - \gamma_{cr})$.

\begin{figure}[ht]
  \begin{tabular}{ccc}
      \includegraphics[width=5.5cm]{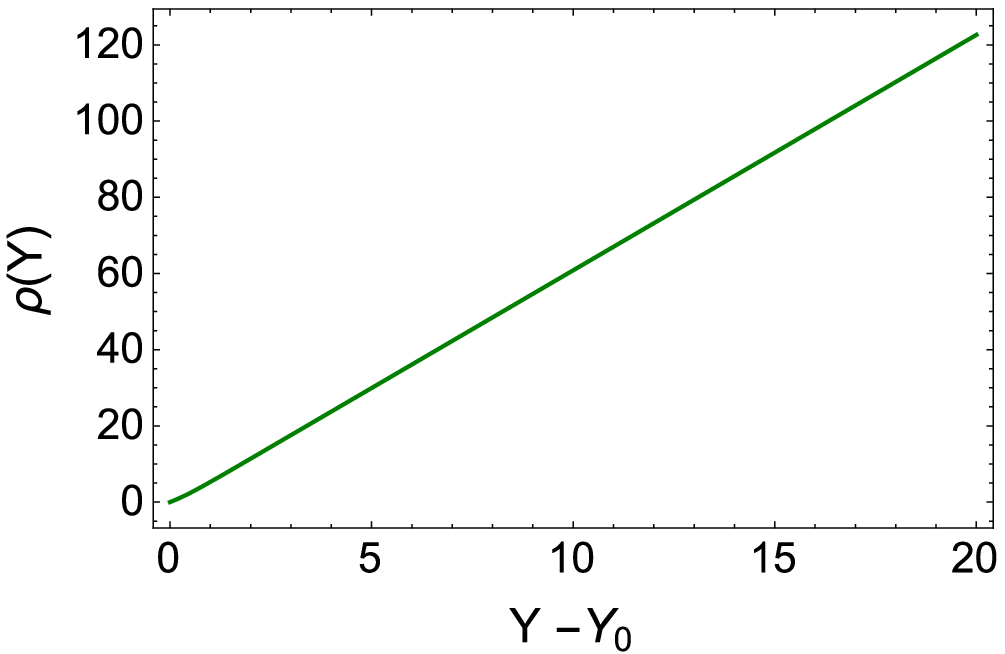} &   \includegraphics[width=5.6cm]{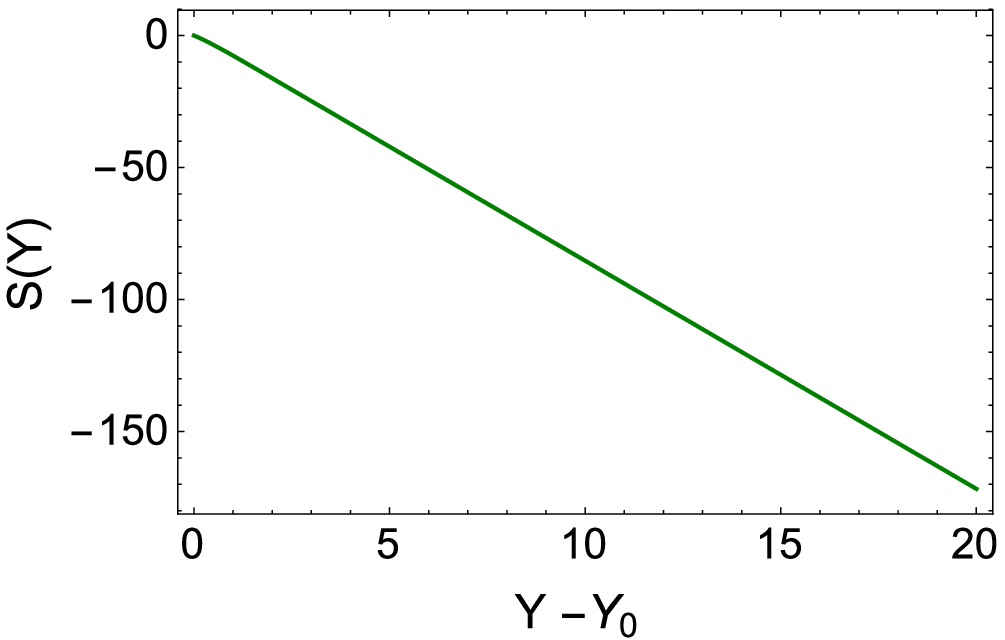} &  \includegraphics[width=5.7cm]{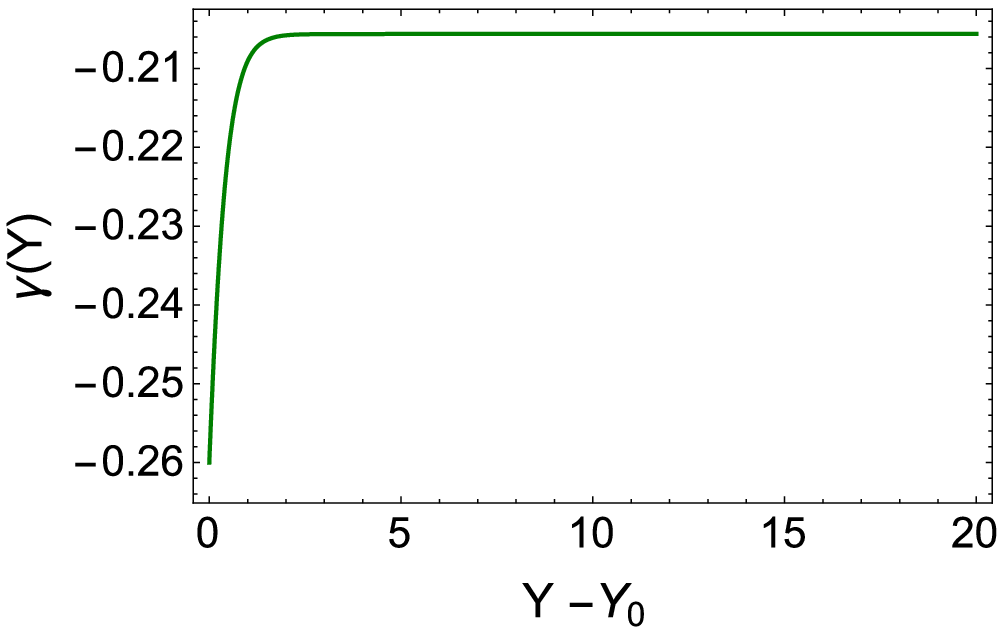}\\
      \fig{sc}-a&\fig{sc}-b & \fig{sc}-c\\
      \end{tabular}
      \caption{The numerical solution to \eq{SCEQ}.
 Constant ${\rm c}$ in \eq{V41} is chosen ${\rm c}$=0.05
 in accord with the description of the HERA data in
 Ref.\cite{CLP}.   $ \as =0.25$. $ \chi\Lb \gamma\Rb=
 2 \psi(1) - \psi(\gamma) - \psi(1 - \gamma)$, where
 $\psi(z) = d \ln \Gamma(z)/d z$ and $\Gamma$ is Euler
  gamma-function. $N_{\rm el}\Lb Y_0, \rho\Rb= {\rm c}
 \Lb Q^2_s\Lb Y_0\Rb/k^2_T\Rb ^{\bar \gamma}$.}
\label{sc}
   \end{figure}

In \fig{sc} we plotted the numerical solutions  of \eq{SCEQ}.  One can see
 that the solution gives $\gamma\Lb Y\Rb$, which approach a constant at 
large
 $Y- Y_0$. This feature stem from \eq{SCEQ}-4 since $N_{el} \to 0$ at large
 $\rho$.

\begin{figure}[ht]
  \begin{tabular}{ccc}
      \includegraphics[width=6.5cm]{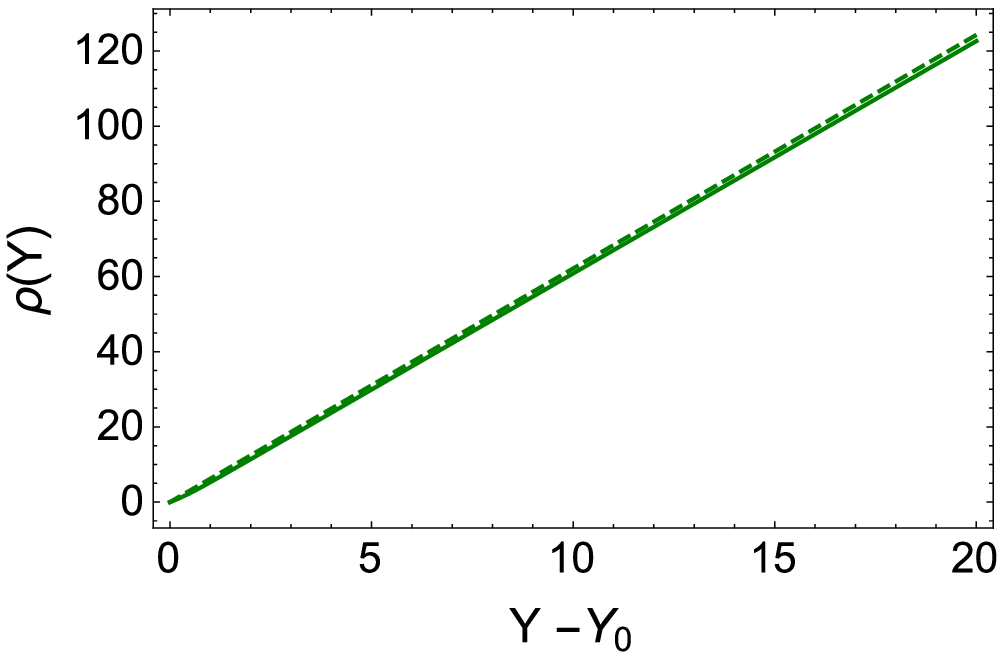} &   ~~~~~~~ &  \includegraphics[width=6.5cm]{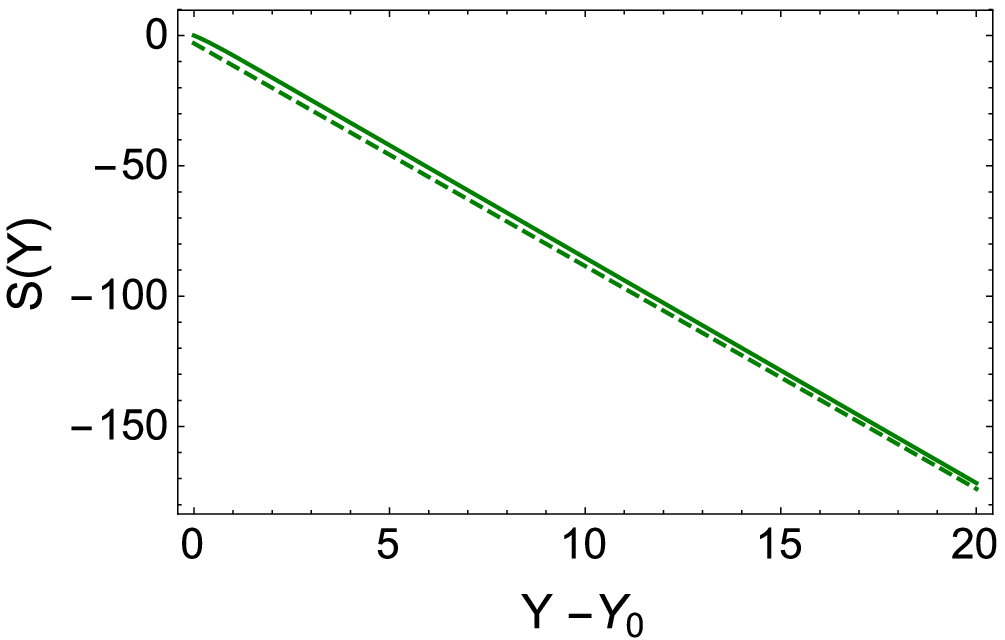}\\
      \fig{sccom}-a&~~~~~~~ & \fig{sccom}-b\\
      \end{tabular}
      \caption{The comparison of the  numerical solution to \eq{SCEQ} and to \eq{V3} in semi-classical approximation.  The solid lines denote the solution to \eq{SCEQ} shown in \fig{sc}, while  the dashed lines describe the solution to \eq{V3} in the semi-classical approximation. For $\xi$ at $Y=0$ we use $\xi $=  0 in the solution to \eq{SCEQ} and 
   $\xi$=3 in the solution to \eq{V3}.   
      Constant ${\rm c}$ in \eq{V41} is chosen ${\rm c}$=0.05.   $ \as =0.25$. $ \chi\Lb \gamma\Rb= 2 \psi(1) - \psi(\gamma) - \psi(1 - \gamma)$, where $\psi(z) = d \ln \Gamma(z)/d z$ and $\Gamma$ is Euler  gamma-function. $N_{\rm el}\Lb Y_0, \rho\Rb= {\rm c} \Lb Q^2_s\Lb Y_0\Rb/k^2_T\Rb ^{\bar \gamma}$.}
\label{sccom}
   \end{figure}

Bearing this in mind one can see that \eq{SC3} and \eq{SCEQ} degenerate
 to \eq{V3} in the semiclassical approach, at large $Y$. Indeed, 
\fig{sccom}
 shows that the solution to \eq{V3} shown in dashed lines in \fig{sccom}, is
 close to the solution of \eq{SC4} for $Y \geq 2$.

Therefore, we need to consider the solution of \eq{CR1} in the
 entire kinematic region where we can neglect the term proportional
 to $\Lb N^D\Rb^2$ in \eq{EQ6}.

  %%%%%%%%%%%%%%%%%%%%%%%%%%%%%%%%%%%%%%%%

   %%%%%%%%%%%%%%%%%%%%%%%%%%%%%%%%%%%%%%%%%%%%%%%%%%%%%%
   
     %%%%%%%%%%%%%%%%%%%%%%%%%%%%%%%%%%%%%%%%

   %%%%%%%%%%%%%%%%%%%%%%%%%%%%%%%%%%%%%%%%%%%%%%%%%%%%%%   
   
   ~
   
\begin{boldmath}
\subsection{Solution in the region where $N_{\rm el} \,\ll\,1$}
\end{boldmath}
%%%%%%%%%%%%%%%%%%%%%%%%%%%%%%%%%%%%%%%%%%%%%%%%%%%%%% 

~
In this section we consider the solution, in the region where
 $N_{\rm el}\Lb Y_0,x_{01},b\Rb$ is a solution to the linear
 BFKL equation, which has the following general form
\beq    \label{SOLSN1}
N_{\rm el}\Lb Y_0,x_{01},b\Rb\,\,=\,\,\int^{\epsilon + i \infty}_{\epsilon + i \infty}\frac{ d \gamma}{ 2\,\pi\,i}\,n_{\rm in}\Lb \gamma, b\Rb \,e^{ \chi\Lb \gamma\Rb\,Y_0\,+\,\Lb \gamma - 1\Rb \xi}
\eeq
where $n_{\rm in}\Lb \gamma, b\Rb$ should be found from the initial 
condition at $Y_0 = 0$. Recall, that $Y_0 \equiv  \bas Y_0$, as we 
have discussed above and $\xi = \ln\Lb 1/(x^2_{10} Q^2_0\Rb$. In 
the region of large $Y_0$, we  take  only the leading
 term in $\chi\Lb\gamma\Rb \xrightarrow{\gamma \ll 1} 1/\gamma$
into account, and
 take   the integral by the method of steepest descent. As the result
 we obtain a solution in the double log approximation (DLA) of
 perturbative  QCD. At large $Y_0$ we can re-write
 $N^2_{\rm el}\Lb Y_0,x_{01},b\Rb$ in the form

\bea \label{SOLSN2}
N^2_{\rm el}\Lb Y_0,x_{01},b\Rb\,\,&=&\,\,\int^{\epsilon + i \infty}_{\epsilon + i \infty}\frac{ d \gamma d \gamma'}{ (2\,\pi\,i)^2}\,n_{\rm in}\Lb \gamma - \gamma', b\Rb \,n_{\rm in}\Lb  \gamma', b\Rb\,e^{ \Lb \chi\Lb \gamma - \gamma'\Rb\,+\,\chi\Lb \gamma'\Rb\Rb\,Y_0\,+\,\Lb \gamma - 2\Rb \xi}\nn\\
\mbox{MOSD  for integration over~~} \gamma'    &\,=\,&\,\,\int^{\epsilon + i \infty}_{\epsilon + i \infty}\frac{ d \gamma }{ 2\,\pi\,i}\,n^2_{\rm in}\Lb \gamma/2, b\Rb \,e^{ 2 \chi\Lb \gamma/2\Rb\,Y_0\,+\,\Lb \gamma - 2\Rb \xi}\eea
where MOSD means the method of steepest descent.

Bearing \eq{SOLSN2} in mind, we obtain the solution to the BFKL
 equation for $N^D$ in the form
\beq \label{SOLSN3}
n^D\,\Lb Y, Y_0,\xi , b\Rb\,\,=\,\,\int^{\epsilon + i \infty}_{\epsilon + i \infty}\frac{ d \gamma }{ 2\,\pi\,i}\,n^D\Lb Y_0, \gamma\Rb\,e^{\chi\Lb \gamma\Rb\Lb Y - Y_0\Rb \,+\,\Lb \gamma - 1\Rb \xi}
\eeq
Calculating  $n^D\Lb Y_0, \gamma\Rb$ from \eq{SOLSN2} we obtain
\beq \label{SOLSN31}
n^D\,\Lb Y, Y_0,\xi , b\Rb\,\,=\,\,\,=\,\int^{\epsilon + i \infty}_{\epsilon + i \infty}\frac{ d \gamma }{ 2\,\pi\,i}\,n^2_{\rm in}\Lb \frac{\gamma +1}{2},b\Rb \,e^{2  \chi\Lb\h(\gamma + 1)\Rb\,Y_0 + \chi\Lb \gamma\Rb\Lb Y - Y_0\Rb \,+\,\Lb \gamma - 1\Rb \xi} 
\eeq
 
\begin{figure}[ht]
  \begin{tabular}{ccc}
      \includegraphics[width=7.5cm]{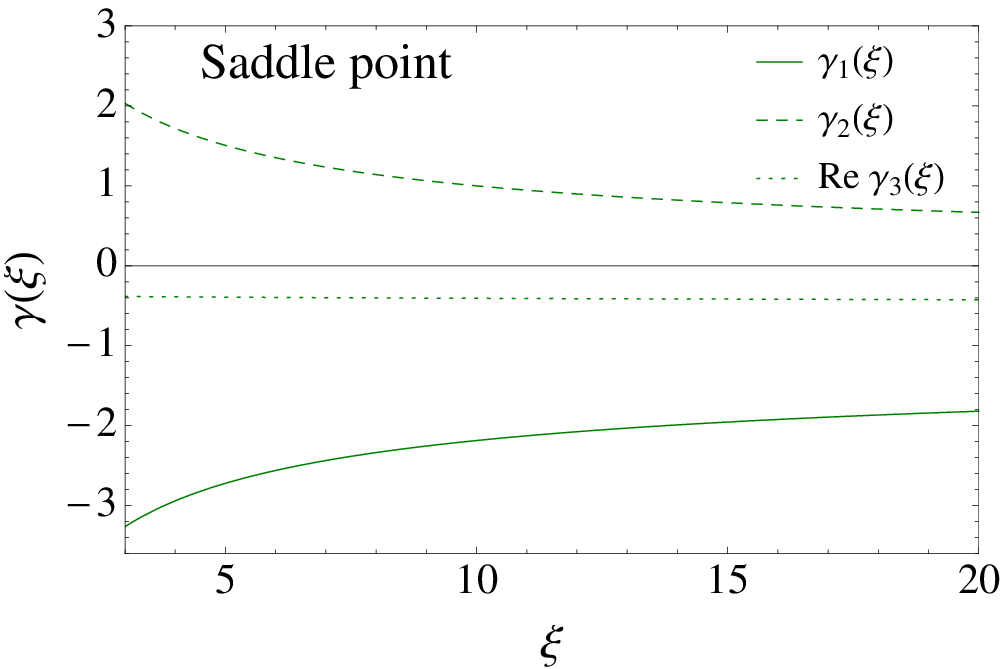} &   ~~~~~~~~~~~ &  \includegraphics[width=7.7cm]{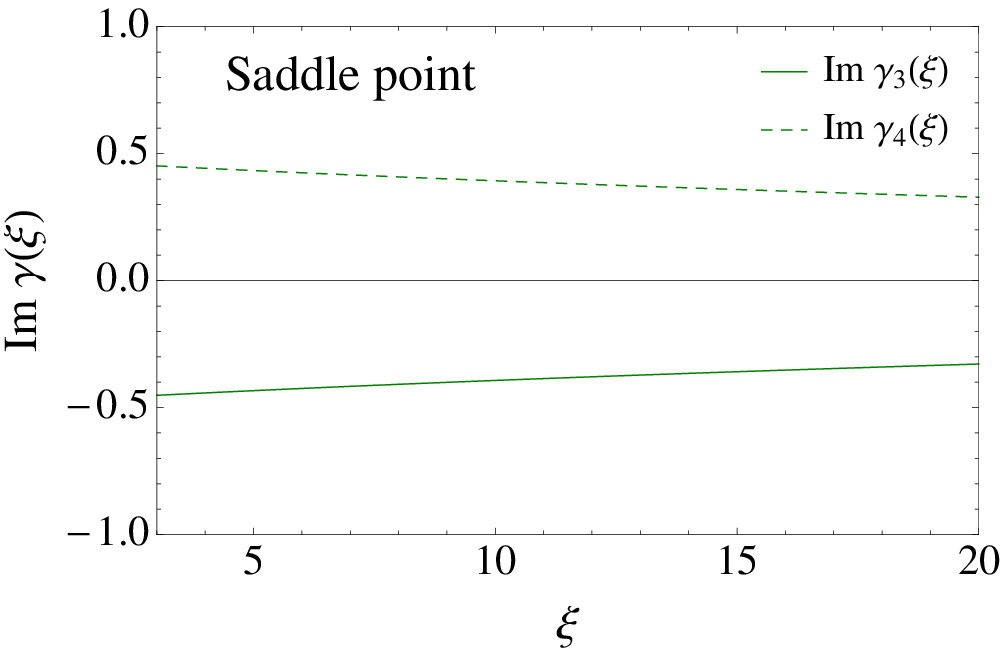}\\
      \fig{sp}-a& & \fig{sp}-b\\
      \end{tabular}
      \caption{$\gamma_{\rm SP}\Lb \xi, Y,Y_0\Rb$ versus  $\xi$ at
 $Y_0$ = 3, $Y$=10.}
\label{sp}
   \end{figure}

  Taking the integral over $\gamma$ in the  DLA,  the equation for the 
saddle
 point takes the form 
  \beq \label{SOLSNSP}
  \xi\,\,=\,\,\frac{Y - Y_0}{ \gamma^2_{\rm SP}}\,\,+\,\,\frac{4 \,Y_0}{\Lb \gamma_{\rm SP} + 1\Rb^2},
  \eeq
  \eq{SOLSNSP} has four solutions (see \fig{sp}).  Two of
 them have  imaginary parts while  other two are real. At $Y - Y_0\, \ll\, \xi - 4 Y_0$ and $\gamma_{\rm SP} \to \sqrt{\frac{Y - Y_0}{\xi\,-\,4\,Y_0 }}$. At $Y - Y_0 \,\ll\, Y_0 \,\ll\,\xi$ one can find: $
\gamma_{\rm SP} \,\,\to\,\,-1 \,+\,\sqrt{\frac{4 Y_0}{\xi - (Y - Y_0)}}$. In
  this   limit we see that our solution satisfies the initial conditions 
in
 the DLA.

For large $Y - Y_0$ the solution has then form

\beq \label{SOLSN4}
N^D_{\rm DLA}\Lb Y, Y_0,\xi , b\Rb\,\,=\,\,\sqrt{\frac{ \pi  ( Y - Y_0 )}{\xi^3}}\,n^2_{\rm in}\Lb\gamma_{SP}= \h \sqrt{\frac{Y -Y_0}{\xi}}, b\Rb\,e^{2 \sqrt{\Lb Y- Y_0\Rb \xi}\,\,-\,\,\,\xi}
\eeq

From \eq{SOLSN4} one can see that the solution reaches the saturation bound at 
\beq \label{SOLSN5}
\xi_{\rm sat}\,=\, 4 \Lb Y -  Y_0\Rb\,~~~~~~~~~~~Q^2_s\Lb Y, Y_0\Rb\,=\,Q^2_0\,e^{ \Lb Y - Y_0\Rb}\,\,=\,\,Q^2_s\Lb Y - Y_0\Rb 
\eeq

and behaves in the vicinity of this bound as
\beq \label{SOLSN6}
n^D_{\rm DLA}\Lb Y, Y_0,\xi , b\Rb\,\,\propto\,\, \Lb x^2_{10}\,Q^2_s\Lb Y   - Y_0\Rb \Rb^{\h}
\eeq

  %%%%%%%%%%%%%%%%%%%%%%%%%%%%%%%%%%%%%%%%

 %%%%%%%%%%%%%%%%%%%%%%%%%%%%%%%%%%%%%%%%%%%%%%    

~

~

~
 %%%%%%%%%%%%%%%%%%%%%%%%%%%%%%%%%%%%%%%%%%%%%%    
\section{The model}   
 %%%%%%%%%%%%%%%%%%%%%%%%%%%%%%%%%%%%%%%%%%%%%% 
As we have discussed, the main idea of building the saturation model has 
been
 formulated in Ref.\cite{IIM}: it is the matching of two analytical solutions
 in the vicinity of the saturation scale, and deep inside of the 
saturation
 domain.

%%%%%%%%%%%%%%%%%%%%%%%%%%%%%%%%%%%%%%%%%%%%%%
 \begin{boldmath}    
\subsection{The input: $N_{el}\Lb r, Y\Rb$ in our saturation model}
\end{boldmath}
 %%%%%%%%%%%%%%%%%%%%%%%%%%%%%%%%%%%%%%%%%%%%%%
  As we have mentioned, the initial condition for the equation for
 $N^D$ (see \eq{EQ7})
 is determined by $N_{el}$, which we have found from the HERA data
 for the deep inelastic structure function in Ref.\cite{CLP}. For
 completeness of presentation we describe the main formulae of this
 model which  illustrates   our procedure for the model building.
 
 In the vicinity of the saturation scale or, in other words,  for
 the dipole size $r$ in the region: $\tau\,\equiv\,r^2\,Q^2_s\Lb Y_0, b\Rb\,\,\to\,\,1$ we use the CGC formula for $N_{el}\Lb r, Y_0\Rb$\cite{KOLEB,IIML,MUT}
 \beq \label{M1}
  N^{\tau \to 1}_{el}\Lb r, Y_0\Rb \,\,=\,\,N_0\,\Lb r^2\,\,Q^2_s\Lb Y_0\Rb\Rb^{1 - \gamma_{cr}}
  \eeq 
 where $N_0$ is the phenomenological parameter that has been found
 in Ref.\cite{CLP}.
 
 $Q_s$ is the saturation momentum which we will discuss below.
 The values of $\gamma_{cr}$ can be found from the following equation:
 \beq \label{M2}
\frac{\chi\Lb \gamma_{cr}\Rb}{1 -   \gamma_{cr}}\,\,=\,\Big{|} \frac{d \chi\Lb \gamma_{cr}\Rb}{d \gamma_{cr}}\Big{|}
\eeq
In \eq{M2}  $\chi\Lb \gamma\Rb$ is the BFKL kernel that takes the form
\beq \label{M3}
\chi\Lb \gamma \Rb\,=\,2\,\psi\Lb 1 \Rb \,-\, \psi\Lb \gamma\Rb \,-\, \psi\Lb 1 -  \gamma\Rb
\eeq
where $\psi\Lb z \Rb$ is the digamma function.

Deep inside of the saturation domain, where
 $\tau\,\equiv\,r^2\,Q^2_s\Lb Y,b\Rb \,\gg
\,1$, we use the analytical solution to the non-linear
 equation given in Ref.\cite{LETU} 
\beq \label{M4}
 N^{\tau \gg 1}_{el}\Lb r, Y_0\Rb\,\,=\,1\,\,-\,\,2\,A\,\exp\Lb - \frac{z^2}{2\,\lambda}\Rb
 \eeq
 where 
 \beq \label{Z1}
z\,\,=\,\,\ln\Lb r^2\,Q^2_s\Lb Y,b\Rb \Rb\,\,=\,\,\ln\Lb r^2 Q^2\Lb Y=Y_{in},   b\Rb\Rb\,\,+\,\,\lambda \Lb Y - Y_{in}\Rb\,\,=\,\,\xi\,\,+\,\,\lambda\,\Lb Y - Y_0\Rb
\eeq
where  $\xi$ and $\lambda$ are related to the behaviour of the saturation 
scale
\beq \label{QS}
Q^2_s\Lb Y_0, b \Rb\,\,=\,\,Q^2\Lb Y = Y_{in}, b\Rb e^{\lambda\,\Lb Y_0 - Y_{in}\Rb}\,\,=\,\,Q^2\Lb Y = Y_{in}, b\Rb \Lb \frac{x_{in}}{x}\Rb^\lambda
\eeq
where $Y_{in} = \ln(1/x_{in})$ shows the initial value of $x$ from 
which we start low $x$  evolution.   This is a phenomenological
 parameter of the model. The phenomenological dependence on $b$
 of the initial saturation scale $Q^2\Lb Y = Y_{in}, b\Rb$ we
 have discussed in the introduction (see \eq{QSB}).  Finally,
 we use the following $Q^2\Lb Y = Y_{in}, b\Rb$
\beq \label{QSIN}
Q^2\Lb Y = Y_{in}, b\Rb\,\,=\,\,Q^2_0\,S\Lb b \Rb\,\,=\,\,Q^2_0\,\Lb m\,b \,K_1\Lb m\,b\Rb\Rb^{\frac{1}{1 - \gamma_{cr}}}
\eeq  
Using \eq{Z}-\eq{QSIN} one can see that
$\xi\,=\,\,\ln\Lb r^2\,Q^2\Lb Y_{in}; b\Rb\Rb$ in \eq{Z}.
 The value of $\lambda$ can be calculated and it is equal
 to $\lambda\,= \,\bas\,\chi\Lb \gamma_{cr}\Rb/(1 - \gamma_{cr})$.
However, in  describing the experimental data in Ref.\cite{CLP}, 
we consider  $\lambda$ as the independent fitting parameter, 
since the next-to-leading correction turns out to be large.

Parameter $A$ in \eq{M4} should be found from the matching procedure 
 of two solution at $z = z_m$:
 
 \beq \label{MC}
 N^{\tau \to 1}\Lb z = z_m \Rb \,=\, N^{\tau \gg  1}\Lb z = z_m \Rb;~~~~~~~~~~~
  \frac{d N^{\tau\,\to\,1}\Lb z = z_m \Rb}{d z_m} \,=\, \frac{d N^{\tau \gg  1}\Lb z = z_m \Rb}{ d z_m}; 
 \eeq 
 
 However,  it turns out that \eq{M4} cannot satisfy \eq{MC}.
 In Ref.\cite{CLM} the correction to \eq{M4} has been found.
 The amplitude of \eq{M4} with the corrections takes the form:
 
 \bea \label{MF}
N^{\tau \,\gg\,1} \Lb z \Rb &\,=\,&1 - 2\,A e^{- {\cal Z}}\,\,-\,\sqrt{2 \lambda} \,A^2\,\frac{1}{\sqrt{{\cal Z} }}e^{- 2 {\cal Z}}\,\,+\,\,{\cal O}\Lb e^{-3 {\cal Z}}\Rb\\
{\cal Z}&=&\frac{\Lb z -  \h A \sqrt{\lambda \pi/2} -
 2\psi(1)\Rb^2}{2 \lambda}\nn
\eea 

which has been used in the \eq{MC}.

It should be stressed that all phenomenological  parameters
 for the elastic amplitude has been extracted from the
 experimental data for $F_2$ at HERA (see Table 1).
%%%%%%%%%%%%%%%%%%%%%%%%%%%%%%%%%%%%%%%%%%%%%%%

\begin{table}[ht]
\begin{tabular}{||l|l|l|l|l|l|l|l|l||}
\hline
\hline
$\lambda $ & $N_0$ & m ($GeV$)& $Q^2_0$ ($GeV^2$) & $m_u $(MeV) &  $m_d $(MeV) &  $m_s $(MeV)&  $m_c $(GeV)  &$\chi^2/d.o.f.$ \\
\hline
0.197& 0.34 & 0.75 & 0.145& 2.3 &4.8& 95&1.4& 178/155 =1.15\\
\hline
0.184 & 0.46 & 0.75 & 0.118& 140 &140 &140& 1.4 & 176/154 = 1.14\\
\hline\hline
\end{tabular}
\caption{Parameters of the model which has been 
extracted from DIS experiment in Ref.\cite{CLP}.   
 $\lambda $, $N_0$,  $m$ and  $Q^2_0$ ($Q^2_0 \,=\,m^2\, x^\lambda_{in}$)
 are fitted parameters. The masses of quarks are not considered as fitted
 parameters and two sets of parameters, that  are shown in the table, relate
  to two choices of the quark masses: the current masses and the masses of
 light quarks  are equal to  $140 \,MeV$ which is the typical infra-red
 cutoff in our approach. }
\label{t1}
\end{table}
%%%%%%%%%%%%%%%%%%%%%%%%%%%%%%%%%%%%%%%%%%%%%%%%%%

%%%%%%%%%%%%%%%%%%%%%%%%%%%%%%%%%%%%%%%%%%%%%%%

%%%%%%%%%%%%%%%%%%%%%%%%%%%%%%%%%%%%%%%%%%%%%%
 \begin{boldmath}    
\subsection{ $N^D\Lb r, Y,Y_0\Rb$ in the model: matching procedure}

\subsubsection{$\tau_0\, = \, r^2\,Q^2_s\Lb Y_0,b'\Rb \,\to\,1$}

\end{boldmath}
 %%%%%%%%%%%%%%%%%%%%%%%%%%%%%%%%%%%%%%%%%%%%%

In the kinematic region where $\tau_0\, = \, r^2\,Q^2_s\Lb Y_0,b'\Rb \,\to\,1$
 , $  \tau\,=\,r^2\,Q^2_s\Lb Y,b\Rb \,\to\,1$  and $\tau_D \,=\,r^2\,Q^2_s\Lb
 Y-Y_0,\vec{b} - \vec{b}' \Rb \,\leq \,1$ we suggest to use \eq{V6} which we
 re-write as follows

\bea \label{MD1}
&&n^D_{\tau_0 \to 1, \tau \to 1,\tau_D \leq 1}\Lb Y, Y_0,\xi , b \Rb\,\,=\\
&&\,\,2 \bar{\gamma}\, \lambda\,\Bigg\{\Bigg( \int d^2 b'\,
\Lb \frac{Q^2_s\Lb Y_0, b'\Rb}{Q^2_0}\Rb^{2\bar{\gamma}} N_1\,\Lb Y - Y_0\Rb \Lb r^2 \,Q^2_s\Lb Y - Y_0, \vec{b} - \vec{b}'\Rb\Rb^{\bar{\gamma}}\Bigg)\,\,+\,\,e^{ - \lambda_1\Lb Y - Y_0\Rb} N^2_{el}\Lb Y, r, b\Rb\Bigg\}\nn\eea
  
  The first term in \eq{MD1}   corresponds to the first term in \eq{V6}, 
which
 we simplify taking into account the experience in the description of the
 elastic amplitude (see \cite{ IIM,SATMOD0,SATMOD1,BKL,SATMOD2,SATMOD3,
SATMOD4,SATMOD5,SATMOD6,SATMOD7,SATMOD8,SATMOD9,SATMOD10, SATMOD11,
SATMOD12,SATMOD13,SATMOD14,SATMOD15,SATMOD16,SATMOD17,CLP,CLP1}). It was
 shown in these papers that  we can describe the solution to the evolution
 equation taking the amplitude in  the form of \eq{M1} replacing $1 -
 \gamma_{cr}$ in \eq{M1} and in \eq{MD1}
by the following expression
\beq \label{N4}
1 \,-\,\gamma_{cr}\,\,\to\,\,1 \,-\,\gamma_{cr}\,\,-\,\,\frac{1}{2\,\kappa\,\lambda\,Y} \,\ln\ \Lb r^2 \,Q^2_s\Lb b \Rb \Rb
\eeq
where $\lambda = \bas \Lb \chi\Lb \gamma_{cr}\Rb/\Lb 1 - \gamma_{cr}\Rb\Rb$
 and $\kappa \,=\,\chi''\Lb \gamma_{cr}\Rb/\chi'\Lb \gamma_{cr}\Rb$. The 
factor $Y - Y_0$ is introduced to reflect the general features of  the
 first term in \eq{V6} which is proportional to this factor. The saturation
 momentum $Q_s\Lb Y - Y_0, \vec{b} - \vec{b}'\Rb$ in QCD  has the general
 form

\beq \label{MD2}
Q^2_s\Lb Y - Y_0,\vec{b} - \vec{b}'\Rb\,\,=\,\,Q^2_s \Lb Y_0,b'\Rb
 \,e^{\kappa\,\Lb Y - Y_0\Rb}
\eeq
where $Q_s\Lb Y_0,b'\Rb$ is the initial transverse momentum at
 $Y=Y_0$. However, as we have discussed above, we introduce the
 non-perturbative corrections in the behaviour at large impact 
parameter in the $b$-dependence of the saturation scale. Bearing
 this in mind we  use the following parameterization of
 $Q_s\Lb Y - Y_0,\vec{b} - \vec{b}'\Rb$
\beq \label{QSM}
Q^2_s\Lb Y - Y_0, \vec{b} - \vec{b}', \vec{b}'\Rb\,\,=\,\,Q^2_{0}\,S\Lb \vec{b}'\Rb \,e^{ - m_1\, |\vec{b} - \vec{b}'|}\,e^{\lambda\,Y }\eeq
where  parameters $ Q_0$, $\lambda$ and mass $m$ in $S\Lb b \Rb$
(see \eq{QSIN}) have been determined  in our previous paper \cite{CLP}
 from the fit of the elastic data.   The parameter $m_1$ has to be
 extracted from the fit of the diffraction production as well  as
 parameters $N_1$ and $\lambda_1$. In the leading order of perturbative
 QCD $\lambda_1 = \chi\Lb \tilde{\gamma}\Rb$, but as well as for the
 energy behaviour of the saturation momentum, the higher order corrections
 are large, and we view these two parameters: $\lambda$ and $\lambda_1$ as
 the phenomenological parameters which we have to extract from the 
experimental data.

   We need to use the matching procedure analogous to  \eq{MC} 
 for $n^D \,=\, -\, d {\cal N}^D d Y_0$ using \eq{DSR5} in the form
   
   \beq \label{MD2}
     \displaystyle{ {\cal N}^{D}_{\tau_0 \to 1, \tau \gg 1}( Y, Y_0; r)\,\,=\,\,1\,-\,G\Lb\tau_0\,=\, r^2\,Q^2_s\Lb Y_0; b \Rb,Y_0\Rb\,\,e^{-z^{2}/2\kappa}   };~~n^D_{\tau_0 \to 1, \tau \gg 1}( Y, Y_0; r)\,\,=\,\,
     \tilde{G}\Lb \tau_0,Y_0\Rb\,\,e^{-z^{2}\Lb b \Rb/2\kappa}     
     \eeq

   The matching equations take the form:
   \bea \label{MC2}
    n^D_{\tau_0 \to 1, \tau \to 1, \tau_D \leq 1}\Lb \tau_0^{m},\tau^m\,,\tau_D^m\Rb  \,\,&=&\,\,   n^D_{\tau_0 \to 1, \tau \gg 1}\Lb \tau_0^{m},\tau^m, \tau_D^m\Rb\,;\nn\\
   \frac{\partial}{\partial \ln \tau}  n^D_{\tau_0 \to 1, \tau \to 1, \tau_D \leq 1}\Lb \tau_0^{m},\tau^m,\,  \tau^m_D\Rb \,\,&=&\,\,        \frac{\partial}{\partial \ln \tau}  n^D_{\tau_0 \to 1, \tau \gg 1}\Lb \tau_0^{m},\tau^m, \tau_D^m\Rb\,;\nn\\
       \frac{\partial}{\partial \ln \tau_0}  n^D_{\tau_0 \to 1, \tau \to 1, \tau_D \leq 1}\Lb \tau_0^{m},\tau^m, \tau^m_D\Rb \,\,&=&\,\,        \frac{\partial}{\partial \ln \tau_0}  n^D_{\tau_0 \to 1, \tau \gg 1}\Lb \tau_0^{m},\tau^m,\tau^m_D\Rb\,;
   \eea   
   \eq{MC2} allow us to specify  function 
 $\tilde{G}\Lb \tau_0,Y_0\Rb$ in \eq{MD2}.
   
   We re-write \eq{MD1} using $\tau(b)$ to simplify \eq{MC2} in the form:
   
   \bea \label{MD3}
&&n^D_{\tau_0 \to 1, \tau \to 1,\tau_D \leq 1}\Lb Y, Y_0,\xi , b, b' \Rb\,\,=\\
&&\,\,~~2 \bar{\gamma}\, \lambda\,\Bigg\{\Bigg(\int d^2 b' e^{\bar{\gamma}\lambda Y_0}\Bigg(\frac{S^3(b')\,e^{-m_1|\vec{b} - \vec{b}'|}}{S\Lb b\Rb }\Bigg)^{\bar{\gamma}}
 N_1\,\Lb Y - Y_0\Rb \Lb r^2 \,Q^2_s\Lb Y , \vec{b} \Rb\Rb^{\bar{\gamma}}\Bigg)\,\,+\,\,e^{ - \lambda_1\Lb Y - Y_0\Rb} N^2_{el}\Lb Y, r, b\Rb\Bigg\}\nn\\
 &&=\,\,2 \bar{\gamma}\, \lambda\,\Bigg\{ \Bigg(\int d^2 b' e^{\bar{\gamma}\,\lambda \,Y_0}\Bigg(\frac{S^3(b')\,e^{-m_1|\vec{b} - \vec{b}'|}}{S\Lb b\Rb}\Bigg)^{\bar{\gamma}}\, N_1\Lb Y - Y_0\Rb \,e^{\bar{\gamma}\,z\Lb b \Rb}\Bigg)
\,\,+\,\,e^{ - \lambda_1\Lb Y - Y_0\Rb} N^2_{el}\Lb Y, r, b\Rb\Bigg\} \nn\eea   
   
   From \eq{MD3} we see that function $\tilde{G}\Lb \tau_0,Y_0, \vec{b},\vec{b}'\Rb$ can be written as
  \beq \label{MD4}
  \tilde{G}\Lb \tau_0,Y_0, \vec{b}\Rb \,\,=\,\,  2\,\int d^2 b' \,\,\bar{\gamma}\,\lambda\, e^{\bar{\gamma}\,\lambda \,Y_0}\Bigg(\frac{S^3(b')\,e^{-m_1|\vec{b} - \vec{b}'|}}{S\Lb b\Rb}\Bigg)^{\bar{\gamma}}\,N_2
  \eeq   
      \eq{MC2} degenerate to the following matching conditions for
 $\lambda_1 \,\Lb Y - Y_0\Rb\,\gg\,1$:
      
      \beq \label{MC3}
      N_1\, e^{\bar{\gamma}\,z_m}\,\,=\,\,N_2\,\,e^{-z^{2}_m/2\kappa} \,;~~~~~\bar{\gamma} \,N_1\,e^{\bar{\gamma}\,z_m}\,\,=\,\,-\,N_2\, \frac{z_m}{\kappa}\,e^{-z^{2}_m/2\kappa} \,;
      \eeq

      One can see that \eq{MC3} does not have a solutions for $z_m >0$.  
 It has been found in Ref.\cite{CLM} that the solution deep inside of the
 saturation scale has more general form than we used in \eq{MD2}:
    \beq \label{MC4}
    n^D_{\tau_0 \to 1, \tau \gg 1}( z, Y_0 ; b )\,\,=\,\,
     \tilde{G}\Lb \tau_0,Y_0\Rb\,\,e^{- \Lb z\Lb b \Rb - B\Lb b \Rb\Rb^2/2\kappa} 
     \eeq
     where
     \beq \label{MC5}
     B\Lb b \Rb\,=\,\int^\infty_0 d z' n^D_{\tau_0 \to 1, \tau \gg 1}( z', Y_0;  b )  \,+\,  2 \,\psi(1) 
     \eeq
     \eq{MC3} takes the form
     \beq \label{MC6}
           N_1\, e^{\bar{\gamma}\,z_m}\,\,=\,\,N_2\,\,e^{-\Lb z_m - B\Rb^2/2\kappa} \,;~~~~~\bar{\gamma} \,N_1\,e^{\bar{\gamma}\,z_m}\,\,=\,\,\,N_2\, \frac{- z_m +B }{\kappa}\,e^{-\Lb z_m - B\Rb^2/2\kappa} \,;    
           \eeq 
  The solution to \eq{MC6} is
  
  \beq \label{MC7}
  B - z_m = \kappa \bar{\gamma}; ~~~~~~~~~~~~N_ 2= N_1 e^{\bar{\gamma}\,z_m}\Big{/}e^{  - \h  \kappa \bar{\gamma}^2};
  \eeq
     
     We can use \eq{MC5} to determine the value of $B$. However, bearing
 in mind the many simplifications that we have assumed , we decided to
 view $z_m$ as a free parameter, which we will find from the fit of the
 experimental data.

 ~
 
 ~
   
   %%%%%%%%%%%%%%%%%%%%%%%%%%%%%%%%%%%%%%%%%%%%%%
 \begin{boldmath}   
\subsubsection{$\tau_0\, = \, r^2\,Q^2_s\Lb Y_0,b'\Rb \,>\,1$}

\end{boldmath} 
   %%%%%%%%%%%%%%%%%%%%%%%%%%%%%%%%%%%%%%%%%%%%%% 
In this region we use \eq{DSR6},  which for $n^D$ takes the form:
\beq \label{MD5}
    n^D\Lb Y,Y_0; r; \vec{b}\Rb \,\,=\,\,\frac{z_0\Lb b\Rb}{\kappa}  \mbox{C}^{2}\exp\Lb -\frac{z^2_0\Lb b\Rb}{ \,\kappa} \Rb  
    \eeq
where $z\Lb b \Rb$ is defined in \eq{Z}. The elastic amplitude is
 equal to
\beq \label{MD6}
N_{el}\Lb Y, r, b\Rb \,\,=\,\,1\,-\,\Delta_{\rm el}(Y; r,\vec{b})~~\mbox{with}~~
 \Delta_{\rm el}(Y; r,\vec{b})\,\,=\,\,\mbox{C}\,e^{-  \frac{z^2\Lb b \Rb}{2\,\kappa} }   \eeq

For practical purpose we define this region as $z\Lb b\Rb \,\geq\,z_m$,
 where $z_m$ is the matching point for DIS.

~ ~
 
 ~
   
   %%%%%%%%%%%%%%%%%%%%%%%%%%%%%%%%%%%%%%%%%%%%%%
  
\subsubsection{Kinematics and observables}

   %%%%%%%%%%%%%%%%%%%%%%%%%%%%%%%%%%%%%%%%%%%%%% 
   The experimental data for diffractive production in DIS
 (see Ref.\cite{HERADATA})  are presented using the following
 set of the kinematic variables:
   \beq \label{KO1}
   \beta\,=\,\frac{Q^2}{Q^2 + M^2_X};~~~~~ x_{Bj}\,=\,\beta\,x_{\pom};
   \eeq
   where $Q^2$ is the virtuality of the photon and $M_X$ is
 the produced mass.  The set of kinematic variables, that we
 used, has the following relation to \eq{KO1}:
   \beq \label{KO2} 
   Y \,\,=\,\,\ln\Lb \frac{1}{x_{Bj}}\Rb\,;~~~~Y_0\,\,=\,\,\ln\Lb \frac{1}{x_\pom}\Rb\,;~~~~Y - Y_0\,\,=\,\,\ln\Lb \frac{1}{\beta}\Rb\,;
   \eeq
   
   The main formulae that we use to calculate the experimental
 cross sections are given by \eq{EQ1} and \eq{EQ2}. \eq{EQ2}
  can be re-written in the form, which includes the
 integration over impact parameter $\vec{b}'$, in the form 
   
   \beq \label{KO3}
 \sigma_{\rm dipole}^{diff}(r_{\perp},x,x_0)
\,\,=\,\,\,\,\int\,d^2 b\,d^2 b' \,n^D(Y, Y_0, r, \vec{b}, \vec{b}')\,, 
\eeq   

 The expression for $(\Psi^*\Psi)^{\gamma^*} \equiv \Psi_{\gamma^*}\Lb Q,
 r, z\Rb \,\Psi_{\gamma^*}\Lb  Q, r,z\Rb$ in \eq{EQ1} is well known (see 
Ref.\cite{KOLEB} and references therein)
\begin{align}
  (\Psi^*\Psi)_{T}^{\gamma^*} &=
   \frac{2N_c}{\pi}\alpha_{\mathrm{em}}\sum_f e_f^2\left\{\left[z^2+(1-z)^2\right]\epsilon^2 K_1^2(\epsilon r) + m_f^2 K_0^2(\epsilon r)\right\},\label{WFDIST}   
  \\
  (\Psi^*\Psi)_{L}^{\gamma^*}&
  = \frac{8N_c}{\pi}\alpha_{\mathrm{em}} \sum_f e_f^2 Q^2 z^2(1-z)^2 K_0^2(\epsilon r),
\label{WFDISL}
\end{align}
where T(L) denotes the polarization of the photon
 and $f$ is the flavours of the quarks. $\epsilon^2\,\,=\,
\,m^2_f\,\,+\,\,Q^2 z (1 - z)$.

~

\subsection{Description of the HERA data}
Using \eq{MF} we attempted to describe  the combined set of the   inclusive diffractive cross sections measured by H1 and ZEUS collaboration at HERA\cite{HERADATA}. The measured cross sections were expressed in terms of reduced cross sections
, $\sigma_r^{D(4)}$, which is related
to the measured $ep$ cross section by
\begin{eqnarray}
\frac{{\rm d} \sigma^{ep \rightarrow eXp}}{{\rm d} \beta {\rm d} Q^2 {\rm d} \x_\pom{\rm d} t} =
\frac{4\pi\alpha^2}{\beta Q^4} \ \ \left[1-y+\frac{y^2}{2}\right] \ \
\sigma_r^{D(4)}(\beta,Q^2,x_\pom,t) \ .
\label{FIT1}
\end{eqnarray}
In the paper, the table of  $      x_\pom \sigma_r^{D(3)}(\beta,Q^2,x_\pom) = x_\pom\int d t\,\, \sigma_r^{D(4)}(\beta,Q^2,x_\pom,t) $ are presented at different values of $Q, \beta $ and $x_\pom$. This cross section is equal to  $\,\frac{Q^2}{4 \pi^2}\,\sigma^{\rm diff}\Lb Y, Y_0, Q^2\Rb$
where $\sigma^{\rm diff}\Lb Y, Y_0, Q^2\Rb$ is given by \eq{EQ1}.

We view this paper as the next step in building the saturation model based on the CGC approach.  The first
step have been done in Refs.\cite{CLP} where we build the saturation model for the DIS. The parameters that we found from this fit and which are shown in Table 1 we use for the diffractive production as given and we are not going to change them.  The additional parameters that we used to parametrize the diffraction production cross section are $N_1$, $m_1$ and $\lambda_1$. $N_1$ is proportional to $\bas$ which indicates that the typical values of $N_1$ is small. $\lambda_1 \,\,= \,\, \bas\,\chi\Lb \tilde{\gamma}\Rb \,\approx\,3.67 \,\bas$ in the leading order of perturbative QCD. However, we consider this as a fitting parameter since we expect that it will be heavily affected by the next order calculation.  Recall, that the value of $\lambda $ which is equal to $\lambda\,=\,\bas \chi\Lb \gamma_{cr}\Rb/(1 - \gamma_{cr})\,\approx\,4.88\,\bas$  came out $\lambda \,\approx\,0.2$ from the fit and this value is in accord with the next to leading estimates.

\begin{figure}[ht]
  \begin{tabular}{cc}
      \includegraphics[width=10cm]{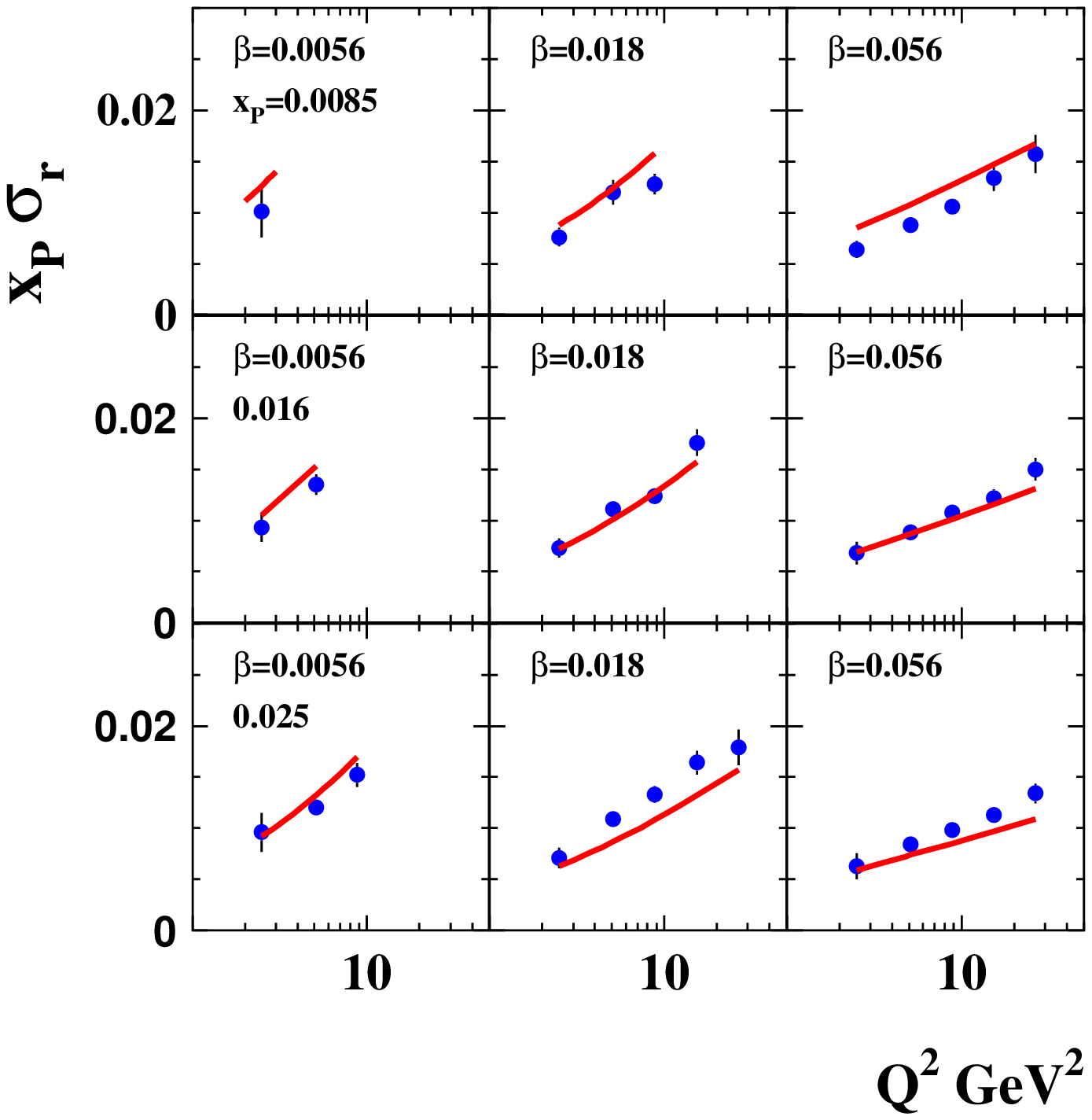} &   \includegraphics[width=10cm]{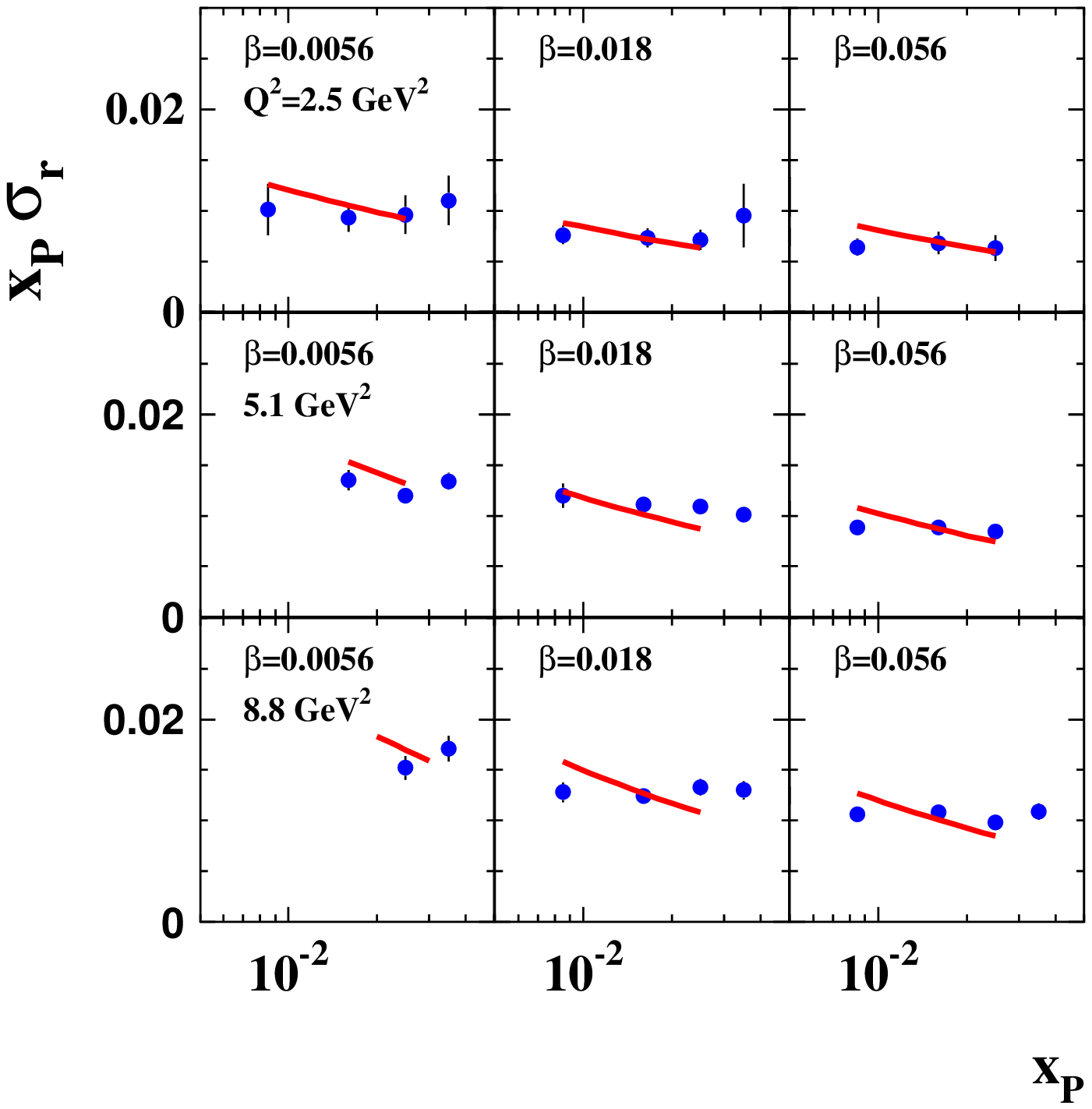}\\
      \fig{f}-a&  \fig{f}-b\\
      \end{tabular}
      \caption{$\sigma^{\rm diff}\Lb Y, Y_0, Q^2\Rb= x_\pom \,\sigma_r$ versus  $Q^2$ at fixed $\beta$ and $x_\pom$ (\fig{f}-a) and  versus  $x_\pom$ at fixed $\beta$ and $Q^2$ (\fig{f}-b). The data are taken from Ref.\cite{HERADATA}. The red curves show the results of the fit with $m_1 = Q_s\Lb Y_0,b\Rb$. The fit in which $m_1$ was a fitted parameter turns out to be so close to this fit that cannot be clearly shown in the picture in spite of the fact that has higher value of $\chi^2/d.o.f.$.}
\label{f}
   \end{figure}

  First, we found a fit 
within  parameters  are equal $N_1 = 7.7\,10^{-4}$, $\lambda_1 = 1.58$ and $m_1 = 2\,GeV$.  
The large value of $m_1$ which describes the non-perturbative behavior of $Q_s\Lb Y - Y_0, \vec{b} - \vec{b}'\Rb$ led us to the idea that even the non-perturbative behavior  of this saturation momentum stems from the CGC physics and determined by $Q_s\Lb Y_0,b\Rb$. Therefore, we fitted the data fixing 
$m_1=Q_s\Lb Y_0,b\Rb$. It turns out that with the  parameters of the fit:  $N_1 = 7.\,10^{-4}$ and $\lambda_1 = 1.48$,  we found the description of the experimental data shown in \fig{f}. Actually, the first fit give the description of the data of the same quality as the second one; and the resulting curves for both fits look the same and cannot be differentiated in the figures.   In spite of the fact that the quality of the fit is not good  we see that \eq{MF} reproduces both $x_\pom$ and $Q$ dependence.

  %%%%%%%%%%%%%%%%%%%%%%%%%%%%%%%%%%%%%%%%

 %%%%%%%%%%%%%%%%%%%%%%%%%%%%%%%%%%%%%%%%%%%%%%    

\begin{figure}[ht]
    \centering
  \leavevmode
      \includegraphics[width=9cm]{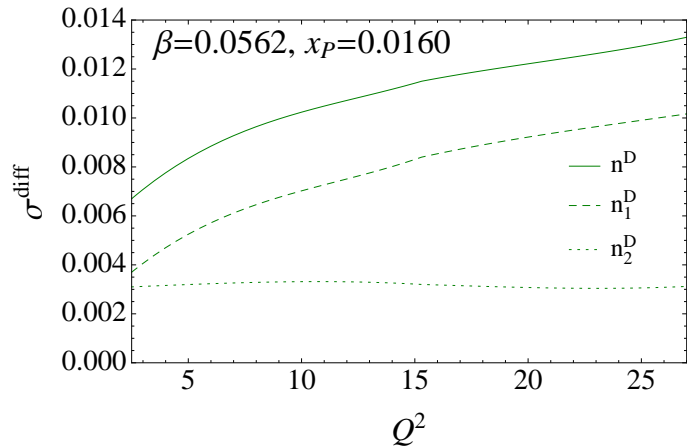}  
      \caption{Contributions of two different sources of the diffractive production: the production of quark-antiquark state ($n^D_2$) and the multi-gluon production ($n^D_1$). }
\label{nd1nd2}
   \end{figure}

We do not expect a good description of the data as we have mentioned. As was expected the values of parameter $\lambda_1$ turns out to be quite different from the leading order estimates in perturbative QCD, which illustrate the need for the next to leading order corrections. We notice that in the experimental kinematic region $\tau_D = r^2 Q^2\Lb Y - Y_0,\vec{b} - \vec{b}'\Rb \,\leq \,1$. Therefore, the most data are in the region 
 which is   outside of the saturation domain.  Hence, \eq{V6} and \eq{asimrepre}, which sums the emission of several gluons in perturbative QCD, has to be tried to describe the data.  The success of the simple model for diffraction production: production of $q \bar{q}$ and $q \bar{q} G$ states\cite{GBKW,GOLEDD,SATMOD0,KOML,MUSCH,MASC,MAR,KLMV}, indicates, that taking into account the emission of several gluons, we will be able to describe the data. We are going to try in a separate further publications.  On the other hand we see that the main contribution stems from the gluon emission. Indeed, in 
 \fig{nd1nd2}  we plot the two different terms of \eq{MF} writing it as $n^D= n^D_1 + n^D_2$ where $n^D_2 \propto N^2_{el}$. One can see that the emission of gluons (the term $n^D_1$) is certainly larger than the contribution of the diffractive production of quark-antiquark state (term $n^D_2$).  
We view this fact as an argument that we have to take into account a large number of emitted gluons.

%%%%%%%%%%%%%%%%%%%%%%%%%%%%%%%%%%%%%%%%%%%%%%%%%%%%%%%%%%%%%%%%%%%%%%%%%%%%%%%%%%%%%%%%%%%%%%%

 %%%%%%%%%%%%%%%%%%%%%%%%%%%%%%%%%%%%%%%%%%%%%%  %%%%%%%%%%%%%%%%%%%%%%%%%%%%%%%%%%%%%%%%%%%%%%%%% 

It should be mentioned that we have some hidden parameters which mostly specify the region of the applicability 
of the perturbative QCD estimates. For example, even in the case of  deep inelastic processes,  we can trust the wave function of perturbative QCD only, at rather large values of $Q^2 \,\geq\,Q^2_0$ with $Q^2_0 \approx 0.7 GeV^2$  since for smaller $Q$ it will be affected by the non-perturbative contributions. As we have mentioned that we consider the matching point $z_m$ of \eq{MC6} as a fitting parameter due to a large uncertainty in  the calculation of $B$ given by \eq{MC5}.  From the fit we specify them as $\beta \,\leq\,0.056$,$x_\pom\,\leq 0.025$, $0.7 \leq Q^2 \leq 27\,GeV^2$.
We found that the fit does not depend on the value of the matching point $z_m$. This confirms that the data  are outside of the saturation region.

    %%%%%%%%%%%%%%%%%%%%%%%%%%%%%%%%%%%%%%%%%% 
  \section{Conclusions}
  %%%%%%%%%%%%%%%%%%%%%%%%%%%%%%%%%%%%%%%%%%
  In the paper we discussed the evolution equations for the diffractive production in the framework of CGC/saturation approach that have been proposed in Ref.\cite{KOLE} and found the analytical solutions in the several kinematic regions. 
  The most impressive features of these  solutions are  that the diffractive production does not show the geometric scaling behaviour being a function of one variable. Even deep in the saturation regions for both diffractive and elastic amplitude the solution turns out to be the product of two functions: one has a geometric scaling behaviour depending on one variable $z$, and the second depends on $z_0$ showing the geometric scaling behaviour in the same way as elastic scattering amplitude at $Y -  Y_0$.
  
  Based on these solutions we suggest a impact parameter dependent saturation model which is suited for the describing the diffraction production both deep in the saturation region and in the vicinity of the saturation scale. Since we are dealing in the diffraction production with two saturation scales:
  $Q\Lb Y, b\Rb$ and $Q\Lb Y - Y_0, \vec{b} - \vec{b}'\Rb$, where $Y = \ln(1/(x_\pom \beta)$ and $Y_0 = \ln(1/\beta)$,  the model includes more information from the theoretical part of the paper than it has been needed for the inclusive DIS.  However, the main key assumptions of the model are the same as for inclusive DIS: the non-perturbative impact parameter behavior is absorbed into two saturation scales.

  Using the model we tried to fit the combined data on diffraction production from H1 and ZEUS collaborations\cite{HERADATA}. We fond that we are able describe both $x_\pom$ and $\beta$ dependence as well as $Q$ behavior of the measured cross sections.  In spite of the sufficiently large $\chi^2/d.o.f.$ we believe that our description give the starting impetus  to find a fit of the experimental data based on the solution of the CGC/saturation equation rather than on describing the diffraction system in smlistic way assuming that onle quark=antiquark pair and one extra gluons are produced. 
  
  Ae the result of the fit we found out that the experimental data are concentrated in the region outside the saturation domain for the produced diffractive system and we intend to try to sum the multi-gluon production in perturbative QCD approach using the formulae which we found in this paper. We are going to publish the result of this kind of approach elsewhere.
  
  We believe that this paper  will revive the interest to the process of the diffractive production which is a unique process  which description needs the
  understanding both the multi-particle generation process and the elastic (diffractive) rescattering at high energy.

    %%%%%%%%%%%%%%%%%%%%%%%%%%%%%%%%%%%%%%%%%% 
  \section{Acknowledgements}
  %%%%%%%%%%%%%%%%%%%%%%%%%%%%%%%%%%%%%%%%%% 
   We thank our colleagues at Tel Aviv university and UTFSM for
 encouraging discussions. Our special thanks go to Asher Gotsman, 
 Alex Kovner and Misha Lublinsky for elucidating discussions on the
 subject of this paper.

  This research was supported by the BSF grant   2012124, by 
   Proyecto Basal FB 0821(Chile),  Fondecyt (Chile) grants  
 1140842, 1170319 and 1180118 and by   CONICYT grant PIA ACT1406. 
\appendix

%%%%%%%%%%%%%%%%%%%%%%%%%%%%%%%%%%%%%%%%%%%%%%%%%%%%%%%%%
\begin{boldmath}
 \section{Solution for $\delta Y \ll 1$ but $\bas \xi \sim 1$. } 
 \end{boldmath}
 %%%%%%%%%%%%%%%%%%%%%%%%%%%%%%%%%%%%%%%%%%%%%%%%%%%%%%%
 
 In this appendix we  obtain the  cross section
 for the diffractive production in the kinematic region, where
 $r^{2}Q^{2}_{s}(Y_{0},b)\approx 1$ and  $\dy\ll1$, but do not use
  the assumption, that $\bas \xi \gg 1$ which we used
 in section IIC-3. As in this section we are dealing with
 the integral
\begin{equation}\label{repre1}
  \displaystyle{\sum_{n=0}^{\infty}(\aal \dy)^{n}  \intee
 \dfrac{1}{\ga ^{n}}\dfrac{\ea}{\ga-\gga}.      }
\end{equation}
We wish to calculate this integral using the special function in
 the most compact and economic way. First, we note that  for $n=0$
 we can take the integral closing the contour of the integration
 over the pole $\gamma = \tilde{\gamma}$.
 For each $n\geq 1$,  we can write the contribution as the convolution integral
 \begin{equation}\label{repreintegral}
\displaystyle{ \intee \dfrac{1}{\ga ^{n}}\dfrac{\ea}{\ga-\gga}}\,\,  =\,\,\displaystyle{e^{\tilde{\gamma}-1} \int_{x}^{1}\dfrac{1}{\Gamma(n)}(-\ln(t))^{n-1} t^{\gga} \dfrac{dt}{t} },
\end{equation}
with $x=e^{-\xi}$.

Plugging   \eqref{repreintegral} into \eqref{repre1} we obtain the following
 expression
\begin{equation}\label{repre2}
n^{D}(Y,Y_{0},\xi,b)\,= \,\,C\Lb Y_0,b\Rb\,e^{(\gga-1)\xi}\left(1+\displaystyle{ \int_{x}^{1}\dfrac{\aal \dy}{\sqrt{\aal \dy \ln(1/t)}}I_{1}(2\sqrt{\aal \dy \ln(1/t)}) t^{\gga}\dfrac{dt}{t}}\right).  
\end{equation}
with
\beq \label{CC}
C\Lb Y_0,b\Rb\,\,=\,\,2 \bas\, \bar{\gamma}\,\kappa\,c^2 \,\Lb \frac{Q^2\Lb Y_0,b\Rb }{Q^2_s\Lb Y_{\rm in}\Rb}\Rb^{2\bar{\gamma}}
\eeq

The next step is to express the integral on the r.h.s. in term of Lommel's
 function. Using that
\begin{equation}\label{entrebessel}
  \dfrac{d}{dt}I_{0}(2\sqrt{\aal \dy \ln(1/t)})=-\dfrac{\aal \dy}{\sqrt{\aal \dy \ln(1/t)}}I_{1}(2\sqrt{\aal \dy \ln(1/t)})\dfrac{1}{t}.
\end{equation} 
 the expression in \eqref{repre2} can be written as follows
\begin{equation}\label{repre3}
n^{D}(Y,Y_{0},\xi,b)\,=\,C\Lb Y_0,b\Rb\,e^{(\gga-1)\xi}\left(1+\displaystyle{ \int_{x}^{1} \dfrac{d}{dt}I_{0}(2\sqrt{\aal \dy \ln(1/t)}) t^{\gga} dt }\right).
\end{equation}
Taking integration by parts into \eqref{repre3} we obtain
\begin{equation}\label{repre4}
n^{D}(Y,Y_{0},\xi,b)\,=\,C\Lb Y_0,b\Rb\,e^{-\xi} I_{0}(2\sqrt{\aal \dy \xi})+\gga e^{(\gga-1)\xi}  \int_{x}^{1} I_{0}(2\sqrt{\aal \dy \ln(1/t)}) t^{\gga} \dfrac{dt}{t} .
\end{equation}
Using $u=\sqrt{\ln(1/t)/\ln(1/x)}$, the integral in the above expression 
 can be  rewritten as
\begin{equation}\label{repre5}
n^{D}(Y,Y_{0},\xi,b)=e^{-\xi} I_{0}(2\sqrt{\aal \dy \xi})+2\gga \xi e^{-\xi}  \int_{0}^{1} e^{\gga \xi(1- u^{2})}I_{0}(2\sqrt{\aal \dy}u) u du .
\end{equation}
Introducing  $z=2i\sqrt{\aal \dy \xi}$, $w=-2i\gga \xi$ into \eqref{repre5} yields the following  representation
 \begin{equation}\label{repre6}
n^{D}(Y,Y_{0},\xi,b)=\,C\Lb Y_0,b\Rb\,\Bigg(e^{-\xi} I_{0}(2\sqrt{\aal \dy \xi})+ e^{-\xi}\dfrac{1}{i}\left( U_{1}(w,z)-iU_{2}(w,z)  \right)\Bigg),
\end{equation}
where $U_{\nu}(w,z)$ denotes the Lommel function of two variables. The 
series representation
\begin{equation}\label{series}
\begin{array}{rcl}
U_{1}(w,z)&=&i\displaystyle{ \sum_{m=0}^{\infty} \left(- \dfrac{\gga \xi}{\sqrt{\aal \dy \xi}} \right)^{2m+1}I_{2m+1}(2\sqrt{\aal \dy \xi})  }\\
U_{2}(w,z)&=&-\displaystyle{ \sum_{m=0}^{\infty} \left(- \dfrac{\gga \xi}{\sqrt{\aal \dy \xi}} \right)^{2m+2}I_{2m+2}(2\sqrt{\aal \dy \xi})  }
\end{array}
\end{equation}
becomes
 \begin{equation}\label{repre7}
n^{D}(Y,Y_{0},\xi,b)=\,C\Lb Y_0,b\Rb\,\Lb e^{-\xi} I_{0}(2\sqrt{\aal \dy \xi})+ e^{-\xi}\displaystyle{\sum_{m=1}^{\infty}(-1)^{m}\left( -\dfrac{\gga \xi}{\sqrt{\aal \dy \xi}} \right)^{m} I_{m}(2\sqrt{\aal \dy \xi})  }\Rb.
\end{equation}
As $\xi\gg 1$, the expression \eqref{repre7} it is not a suitable
 representation for the asymptotic analysis. This problem is 
resolved considering the generating function of the Bessel 
functions that  can be  written as follows
\begin{equation}\label{genera}
  e^{(z/2)\left( t+t^{-1} \right)}=\displaystyle{ \sum_{m=-\infty}^{\infty}t^{m}I_{m}(z)  }
\end{equation}
Introducing \eq{genera} into \eq{repre7}, and using that 
 $I_{m}(z)=I_{-m}(z)$, we obtain
\beq \label{repre}
n^{D}(Y,Y_{0},\xi,b)=\,C\Lb Y_0,b\Rb\,\Lb e^{(\gga-1) \xi +\dfrac{\aal \dy}{\gga}}-\displaystyle{e^{-\xi} \sum_{m=1}^{\infty}\left( \dfrac{1}{\gga}\sqrt{\dfrac{\aal \dy}{\xi}}  \right)^{m} I_{m}(2\sqrt{\aal \dy \xi}).}\Rb
\eeq
At large $\aal \dy \xi $,  \eq{repre} takes the form
\begin{equation}\label{asimrepre}
n^{D}(Y,Y_{0},\xi,b)=\,C\Lb Y_0,b\Rb\,\Lb e^{(\gga-1) \xi +\dfrac{\aal \dy}{\gga}}-2\sqrt{\pi}\displaystyle{ \left( \dfrac{(\aal \dy)^{3}}{\xi}  \right)^{1/4} \dfrac{e^{-\xi+2\sqrt{\aal \dy\xi}}}{\gga -\sqrt{\aal \dy/\xi}}. }\Rb
\end{equation}
One can see that this equation coincides with \eq{V6} if
 we replace $\chi\Lb \gamma\Rb$ by $1/\gamma$ and consider
$\sqrt{\aal \dy/\xi}$ being much larger than $\gga$.

\end{document}